\documentclass[prx,twocolumn,superscriptaddress,citeautoscript,showpacs,amsart,longbibliography,nofootinbib]{revtex4-1}

\usepackage{graphicx}
\usepackage{multirow}
\usepackage{color}
\usepackage{bm}
\usepackage{times}
\usepackage{amsmath,bm,amsfonts}
\usepackage{dcolumn}
\usepackage{graphicx}
\usepackage{latexsym}

\usepackage{ulem} % to strike things out \normalem % usual emph
\usepackage{braket}

\usepackage{cancel}

\def\ket#1{|#1\rangle }

\def\bb{\mathbb}
\def\Pf{{\rm Pf}}
\def\d{\partial}

\begin{document}
\title{Symmetry representation approach to topological invariants in $C_{2z}T$-symmetric systems
}

\author{Junyeong \surname{Ahn}}
\affiliation{Department of Physics and Astronomy, Seoul National University, Seoul 08826, Korea}

\affiliation{Center for Correlated Electron Systems, Institute for Basic Science (IBS), Seoul 08826, Korea}

\affiliation{Center for Theoretical Physics (CTP), Seoul National University, Seoul 08826, Korea}

\author{Bohm-Jung \surname{Yang}}
\email{bjyang@snu.ac.kr}
\affiliation{Department of Physics and Astronomy, Seoul National University, Seoul 08826, Korea}

\affiliation{Center for Correlated Electron Systems, Institute for Basic Science (IBS), Seoul 08826, Korea}

\affiliation{Center for Theoretical Physics (CTP), Seoul National University, Seoul 08826, Korea}

\date{\today}

\begin{abstract}
We study the homotopy classification of symmetry representations to describe the bulk topological invariants protected by the combined operation of a two-fold rotation $C_{2z}$ and time-reversal $T$ symmetries.
We define topological invariants as obstructions to having smooth Bloch wave functions compatible with a momentum-independent symmetry representation.
When the Bloch wave functions are required to be smooth, the information on the band topology is contained in the symmetry representation.
This implies that the $d$-dimensional homotopy class of the unitary matrix representation of the symmetry operator corresponds to the $d$-dimensional topological invariants.
Here, we prove that the second Stiefel-Whitney number, a two-dimensional topological invariant protected by $C_{2z}T$, is the homotopy invariant that characterizes the second homotopy class of the matrix representation of $C_{2z}T$.
As an application of our result, we show that the three-dimensional bulk topological invariant for the $C_{2z}T$-protected topological crystalline insulator proposed by C. Fang and L. Fu in Phys. Rev. B {\bf 91}, 161105(R) (2015), which we call the 3D strong Stiefel Whitney insulator, is identical to the quantized magnetoelectric polarizability.
The bulk-boundary correspondence associated with the quantized magnetoelectric polarizability shows that the 3D strong Stiefel-Whitney insulator has chiral hinges states as well as 2D massless surface Dirac fermions.
This shows that the 3D strong Stiefel Whitney insulator has the characteristics of both the first order and the second order topological insulators, simultaneously, which is consistent with the recent classification of higher-order topological insulators protected by an order-two symmetry.
\end{abstract}

\maketitle

\section{Introduction}

Topological crystalline insulators (TCIs) are insulators whose bulk properties cannot be adiabatically connected to those of atomic insulators due to symmetry~\cite{hasan2010colloquium,qi2011topological,chiu2016classification,ando2015topological}. 
Accordingly, the topological invariant of a TCI can generally be considered as an obstruction to finding exponentially localized symmetric Wannier functions.
Since the construction of symmetric Wannier functions requires both the exponential localization of the wave functions and the invariance under symmetry, 
the topological invariant of a TCI can be defined in two different ways. First, when the symmetry representation is trivial in momentum space --- meaning that it can be induced by a symmetry representation of Wannier functions in real space, the topological invariant of a TCI becomes an obstruction to having smooth wave functions in momentum space~\footnote{Topological obstruction to smoothness is due to the obstruction to continuity, because topology is about continuity. In practice, the obstruction to smoothness is identical to the obstruction to continuity because the Bloch wave functions are smooth in topologically trivial phases. Let us note that a nontrivial band topology does not induce Bloch wave functions that are continuous but are not smooth.}.
On the other hand, when the wave functions are assumed to be smooth in the Brillouin zone, the topological invariant of a TCI is encoded in the matrix representation of the symmetry operator, the so-called sewing matrix, and appears as an obstruction to finding a trivial sewing matrix in the Brillouin zone.
Although both approaches eventually lead to the same classification of TCIs, the different ways of defining a topological invariant provide complementary views of understanding the nature of the relevant TCIs. Identifying the topological obstruction under symmetry constraints and the nature of the associated topological invariants, taking into account all possible space groups and magnetic space groups in crystals, is definitely one central issue in the study of TCIs.

Recently, it has been shown that a 2D system with a magnetic symmetry under $C_{2z}T$ can be characterized by the first Stiefel-Whitney number $w_1$ and the second Stiefel-Whitney number $w_2$~\cite{shiozaki2017topological,ahn2018band}.
As $C_{2z}T$ is a local symmetry operator in momentum space, and also an antiunitary symmetry operator that satisfies $(C_{2z}T)^{2}=1$,  its constant representation can be taken to be the identity matrix: $C_{2z}T\ket{u_{n\bf k}}=\ket{u_{n\bf k}}$~\cite{fang2015new}.
This gauge choice is called a real gauge because the residual gauge transformation belongs to a real unitary group, i.e., an orthogonal group.
As the symmetry representation is constant in this gauge, topological invariants are defined as obstructions to the smoothness of Bloch wave functions.
The relevant 1D and 2D topological invariants are $w_1$ and $w_2$.
$w_1$ is equivalent to the quantized Berry phase, and $w_2$ is identical to the $Z_2$ monopole charge~\cite{morimoto2014weyl,fang2015topological,zhao2017pt} characterizing a nodal line semimetal in $PT$-symmetric 3D spinless fermion systems~\cite{ahn2018band}.
In terms of more physical concepts, $w_1$ is the quantized electric dipole moment, and $w_2$ is the quantized electric quadrupole moment~\cite{ahn2018band}.
In crystals, the electric quadrupole moment is not well-defined when the electric dipole moment is nontrivial~\cite{benalcazar2017quantized}.
Similarly, $w_2$ becomes a well-defined 2D topological invariant of an insulator only when $w_1$ is trivial~\cite{ahn2018band}, and the insulator with $w_2=1$ was dubbed a 2D Stiefel-Whitney insulator (SWI)~\cite{ahn2018band}.

The 2D SWI is a fragile topological insulator~\cite{po2018fragile}, which means that its Wannier obstruction is fragile against adding trivial bands (i.e., bands with $w_1=w_2=0$): it has a Wannier obstruction when the number of occupied bands is two, but the obstruction disappears when an additional trivial band is added. 
After the Wannier obstruction disappears, the resulting atomic insulator with $w_2=1$ still carries a nonzero electric quadrupole moment originating from the Wannier centers residing on the boundary of the unit cell, so it corresponds to an obstructed atomic insulator~\cite{cano2018topology,bouhon2018wilson,wang2018higher,bradlyn2018disconnected,ahn2019failure}.
The fragile nature of the 2D SWI is also reflected in the Whitney sum formula for Stiefel-Whitney numbers, which is an algebraic rule for adding $w_1$ and $w_2$ of the subbands below the Fermi level~\cite{ahn2019failure}.

In this paper, we revisit the $C_{2z}T$-protected topological invariants in the perspective of the homotopy classification of symmetry representations in momentum space~\cite{wang2010equivalent,hughes2011inversion,bradlyn2017topological,alexandradinata2018no}.
When we choose a smooth gauge instead of a real gauge, a nontrivial topology should manifest as an obstruction to taking a constant symmetry representation. Accordingly, in principle, a homotopy classification of the corresponding sewing matrix should give a classification of topological phases~\cite{bradlyn2017topological}.
Here, we establish the relationship between the representation of $C_{2z}T$ symmetry and the second Stiefel-Whitney number.
To this end, we define momentum-independent symmetry representations as a trivial symmetry representation, trivial symmetry representations.
This definition allows a finer classification than the case based on the Wannier obstruction, because even obstructed atomic insulators, which are all trivial in terms of the Wannier obstruction, can be distinguished in this way.
In real space, trivial insulators in our definition correspond to the atomic insulators whose Wannier centers can be adiabatically deformed to the center of the unit cell.
Although this definition of topological triviality depends on the choice of the unit cell, it is useful because it provdies a {\it relative} classification of the obstructed atomic insulators. Given this definition, we can use the usual homotopy group  to classify topological insulators because a constant map forms the trivial element of the homotopy group.
Explicitly, we find that the $d$-th homotopy class of the sewing matrix directly gives the conventional $d$-dimensional topological invariants, that is, the Stiefel-Whitney numbers $w_1$ and $w_2$ for $d=1$ and $d=2$, and the magnetoelectric polarizability $P_3$ for $d=3$.
Let us note that although the relation between the sewing matrices and topological invariants has already been shown for $w_1$ and $P_3$ by using the Berry connection and curvature, such a relation for $w_2$ has not been known yet.
A merit of this approach based on the homotopy class of the sewing matrix is that it provides the Whitney sum formula for Stiefel-Whitney numbers in a more comprehensive way as compared to the formulation in a real gauge as shown below.

%%%%%%%%%%%%%%%%%%%%%%%%%%%%%%%%%%%%%%%%   FIGURE   %%%%%%%%%%%%%%%%%%%%%%%%%%%%%%%%%%%%%%%%%%%%%%%%%%%%%%
\begin{figure}[b!]
\includegraphics[width=8.5cm]{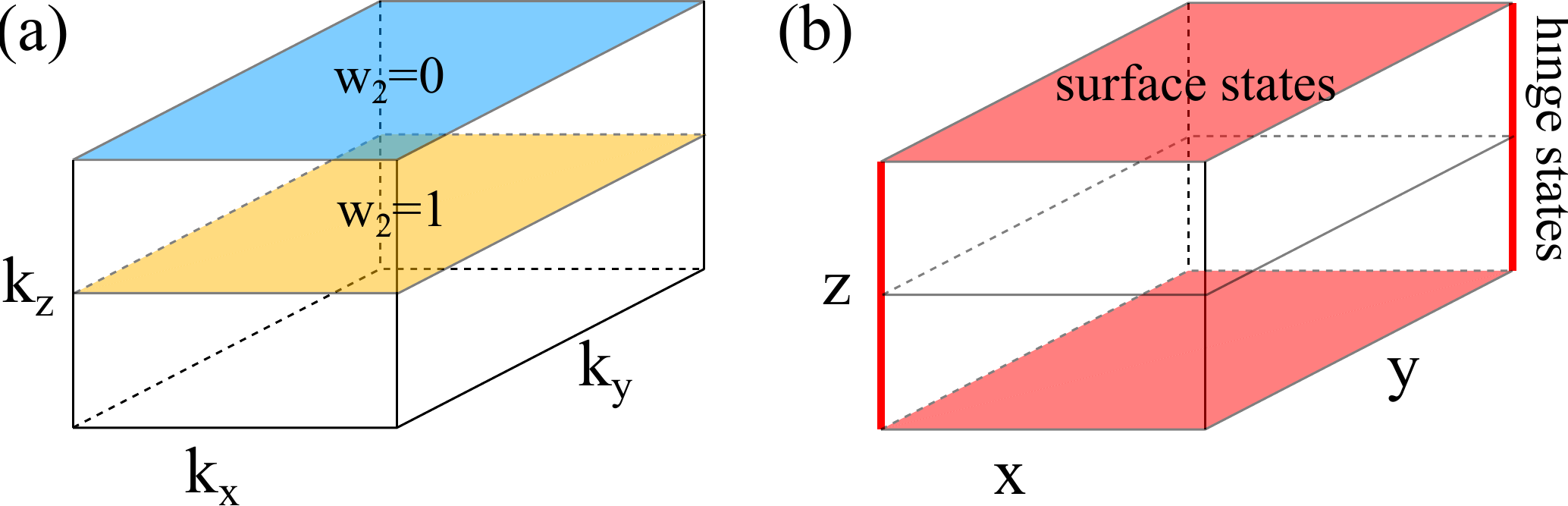}
\caption{3D strong Stiefel-Whitney insulator (SWI) protected by $C_{2z}T$ symmetry.
(a) Schematic figure describing the second Stiefel-Whitney numbers on the $C_{2z}T$-invariant planes in momentum space.
In a 3D strong SWI, $w_2(k_z=\pi)-w(k_z=0)=1$ modulo two.
(b) Schematic figure describing the gapless states on the surface and hinges in real space.
An odd number of 2D Dirac fermions appear on each of the top and bottom surfaces.
1D chiral fermions appear on the side hinges.
}
\label{fig.scheme}
\end{figure}
%%%%%%%%%%%%%%%%%%%%%%%%%%%%%%%%%%%%%%%%   FIGURE   %%%%%%%%%%%%%%%%%%%%%%%%%%%%%%%%%%%%%%%%%%%%%%%%%%%%%%

As an application of our approach, we derive the relation between the second Stiefel-Whitney number and the higher-order band topology of the three-dimensional topological insulator with $C_{2z}T$ symmetry proposed in Ref.~\onlinecite{fang2015new}.
In 3D momentum space, there are two $C_{2z}T$-invariant planes with $k_z=0$ and $k_{z}=\pi$, and the corresponding second Stiefel-Whitney invariants can be written as $w_{2}(0)$ and $w_{2}(\pi)$, respectively.
Thus, a 3D $Z_{2}$ topological invariant $\Delta_{3D}$ can be defined as $\Delta_{3D}\equiv w_{2}(\pi)-w_{2}(0)$~\cite{fang2015new}.
Since $\Delta_{3D}$ originates from $w_{2}$ in $C_{2z}T$-invariant planes, we call the 3D topological insulator with $\Delta_{3D}=1$ as {\it a 3D strong Stiefel-Whitney insulator}.
The 3D strong SWI was originally proposed to have only Dirac surface states as anomalous boundary states on the two $C_{2z}T$-invariant surfaces, which are normal to the $z$ direction~\cite{fang2015new}.
However, recent classifications~\cite{geier2018second,trifunovic2019higher} of higher-order topological insulators~\cite{
benalcazar2017quantized,
benalcazar2017electric,
serra2018observation,imhof2018topolectrical,
langbehn2017reflection,song2017d,
geier2018second,trifunovic2019higher,
schindler2018bismuth,
matsugatani2018connecting, %18.04
wang2018higher, %18.06
franca2018anomalous, %18.07
calugaru2018higher, %18.08
benalcazar2018quantization, %18.09
ezawa2018higher, %17.09
%ezawa2018minimal, %18.01
%ezawa2018topological, %18.06
ezawa2018simple, %18.07
zhang2013surface,fang2017rotation,khalaf2018higher,
kooi2018inversion,
varnava2018surfaces,vanmiert2018higher,
yue2019symmetry,schindler2018higher,
ezawa2018strong,ezawa2018magnetic} indicate that the 3D strong SWI has additional anomalous chiral states on side hinges, so-called chiral hinge states [Fig.~\ref{fig.scheme}].
That is, the 3D strong SWI is a mixed-order topological insulator, because the $n$th-order topological insulator is defined by the presence of $(d-n)$-dimensional anomalous boundary states.
Below we show that the mixed-order band topology in this system results from the nontrivial magnetoelectric polarizability $P_3$ of the bulk, which satisfies $\Delta_{3D}=2P_3$. This relation is derived based on the fact that the 3D topological invariant corresponds to the third homotopy class of the sewing matrix.

This paper is organized as follows.
In Sec.~\ref{sec.homotopy}, we analyze the general properties of the sewing matrix for $C_{2z}T$ symmetry.
Our main technical tool is the exact sequences of homotopy groups that have been used to classify the space of Hamiltonians for topological insulators~\cite{turner2012quantized,trifunovic2017bott,sun2018conversion}.
Using the exact sequences, in one and two dimensions, we show that the homotopy class of the sewing matrix gives the same classification of topological phases as the homotopy class of the Hamiltonian spaces.
Then, the connection between the second and third homotopy classes of the sewing matrix is shown.
These general results are elaborated further in the following sections.
In Secs.~\ref{sec.first} and~\ref{sec.second}, we show the explicit relations between the sewing matrix in a smooth gauge and the transition functions in a real gauge.
The first and second Stiefel-Whitney numbers defined in a real gauge are matched to the first and second homotopy classes of the sewing matrix in a smooth gauge.
In 3D, we first determine the condition for $\Delta_{3D}$ to be a well-defined topological invariant in Sec.~\ref{sec.strong}.
After that, we show that the bulk magnetoelectric polarizability is determined by the second Stiefel-Whitney numbers of $C_{2z}T$-invariant planes when $\Delta_{3D}$ is well-defined in Sec.~\ref{sec.third}.
We review the bulk-boundary correspondence between the anomalous boundary states and the bulk magnetoelectric polarizability in Sec.~\ref{sec.bulk-boundary},
and demonstrate, by using a tight-binding toy model, that the 3D topological insulator with $\Delta_{3D}=1$ has both anomalous Dirac surface states and chiral hinge states in Sec.~\ref{sec.TB}.
Finally, we discuss some generalizations of our results in Sec.~\ref{sec.discussion}.

\section{Homotopy groups of the sewing matrix}
\label{sec.homotopy}

Let us begin by studying general aspects of the homotopy groups of the sewing matrix $G$ for $C_{2z}T$.
$G$ is defined as
\begin{align}
\label{eq.sewing}
G_{mn}({\bf k})
&=\braket{u_{m(-C_{2z}{\bf k})}|C_{2z}T|u_{n\bf k}},
\end{align}
where $-C_{2z}{\bf k}=(k_x,k_y,-k_z)$ and $\ket{u_{n\bf k}}$ is the cell-periodic part of the Bloch state.
We are interested in the ground state of the system and study the topology of the occupied states, so hereafter we assume that $m$ and $n$ run over the indices of occupied bands.
Since $(C_{2z}T)^2=(C_{2z})^2T^2=1$ in both spinless and spinful systems, $G$ satisfies
\begin{align}
\label{eq.sewing_constraint}
G_{mn}({\bf k})
&=G_{nm}(-C_{2z}{\bf k}).
\end{align}
Under a gauge transformation
$\ket{u_{n\bf k}}\rightarrow \ket{u'_{n\bf k}}
=U_{mn}({\bf k})\ket{u_{m\bf k}}$,
the sewing matrix transforms as
\begin{align}
\label{eq.sewing_gauge}
G_{mn}({\bf k})
\rightarrow G'_{mn}({\bf k})
=[U^{\dagger}( -C_{2z}{\bf k})G({\bf k})U^*({\bf k})]_{mn},
\end{align}
where $G'_{mn}({\bf k})=\braket{u'_{m(-C_{2z}{\bf k})}|C_{2z}T|u'_{n\bf k}}$.
If we choose smooth wave functions for occupied states, the corresponding sewing matrix also becomes smooth.
The nontrivial homotopy class of $G$ characterizes the obstruction to taking a uniform representation $G({\bf k})=G_0$ independent of ${\bf k}$.

On a $C_{2z}T$-invariant plane, either the $k_z=0$ or $k_z=\pi$ plane, $G^T({\bf k})=G({\bf k})$ according to Eq.~\eqref{eq.sewing_constraint}. 
Such a symmetric unitary matrix can be written as 
\begin{align}
\label{eq.sewing=U}
G({\bf k})=U_G({\bf k})U_G^T({\bf k}),
\end{align}
where $U_G$ is a unitary matrix describing a unitary transformation from a smooth gauge to a real gauge.
As a redefinition $U_G({\bf k})\rightarrow O({\bf k})U_G({\bf k})$ for any $O({\bf k})\in O(N)$ does not change $G({\bf k})$, we obtain
\begin{align}
G({\bf k})\in U(N)/O(N),
\end{align}
on $C_{2z}T$-invariant planes, where $N$ denotes the number of occupied bands.

Since a nontrivial homotopy class of $G({\bf k})$ is an obstruction to taking a constant symmetry representation, it classifies possible topological phases for $N$ occupied bands.
To get a well-defined classification of topological phases, however, one should carefully identify the homotopy classes that are related to each other by gauge transformations, because a smooth gauge transformation can change the homotopy class of $G$.
After that, the homotopy classification should be consistent with the classification of the Hamiltonian space.
In fact, we show below that
\begin{align}
\frac{\pi_{d}\left[U(N)/O(N)\right]}{\rm Gauge DOF}
\simeq \pi_{d}\left[\frac{O(N+M)}{O(N)\times O(M)}\right]_{M\rightarrow \infty},
\end{align} 
where $d=1,2$,~\footnote{
We focus on $d=1,2$ here because we consider $C_{2z}T$ symmetry.
However, this equation can be extended to higher-dimensional systems with $PT$ symmetry with $(PT)^2=1$.
In general, it is valid for any $d\ne 4n$ for a positive integer $n$ as shown in Appendix~\ref{sec.homotopy_sequence}.
When $d=4n$ for some positive integer $n$, the band topology is characterized by the $2n$-th Chern class, so the nontrivial band topology does not require $PT$ symmetry and persists without the symmetry.
Therefore, the classification of the sewing matrix does not give the full classification of band topology in the case.}
${\rm GaugeDOF}$ is the image of the map $j^*:\pi_d[U(N)]\rightarrow \pi_d[U(N)/SO(N)]$ that is induced by the projection $j:U(N)\rightarrow U(N)/O(N)$, and $O(N+M)/[O(N)\times O(M)]$ is the classifying space of the real (i.e., $C_{2z}T$-symmetric) Hamiltonians for $N$ occupied and $M$ unoccupied bands~\cite{bzdusek2017robust}.

The above equivalence can be explicitly shown in two steps.
First, we use that 
\begin{align}
\pi_{d}\left[\frac{O(N+M)}{O(N)\times O(M)}\right]_{M\rightarrow \infty}\simeq \pi_{d-1}[O(N)],
\end{align} 
which states that the $d$-dimensional topological phase described by a real Hamiltonian is characterized by the $(d-1)$-th homotopy class of the transition function for real wave functions~\cite{hatcher2002algebraic,hatcher2003vector,fang2015topological}.
Then, we use the equivalence between the formalism in the smooth gauge and that in the real gauge:
\begin{align}
\label{eq.smooth=real}
\frac{\pi_{d}\left[U(N)/O(N)\right]}{\rm Gauge DOF}
&\simeq \pi_{d-1}[O(N)],
\end{align}
where $d=1,2$, which can be derived from the exact sequence of homotopy groups [See Appendix~\ref{sec.homotopy_sequence}].
We demonstrate the relation between the smooth gauge and the real gauge in more detail for $d=1,2$ in the following sections~\ref{sec.first} and \ref{sec.second}, respectively.

\begin{figure}[t!]
\includegraphics[width=8.5cm]{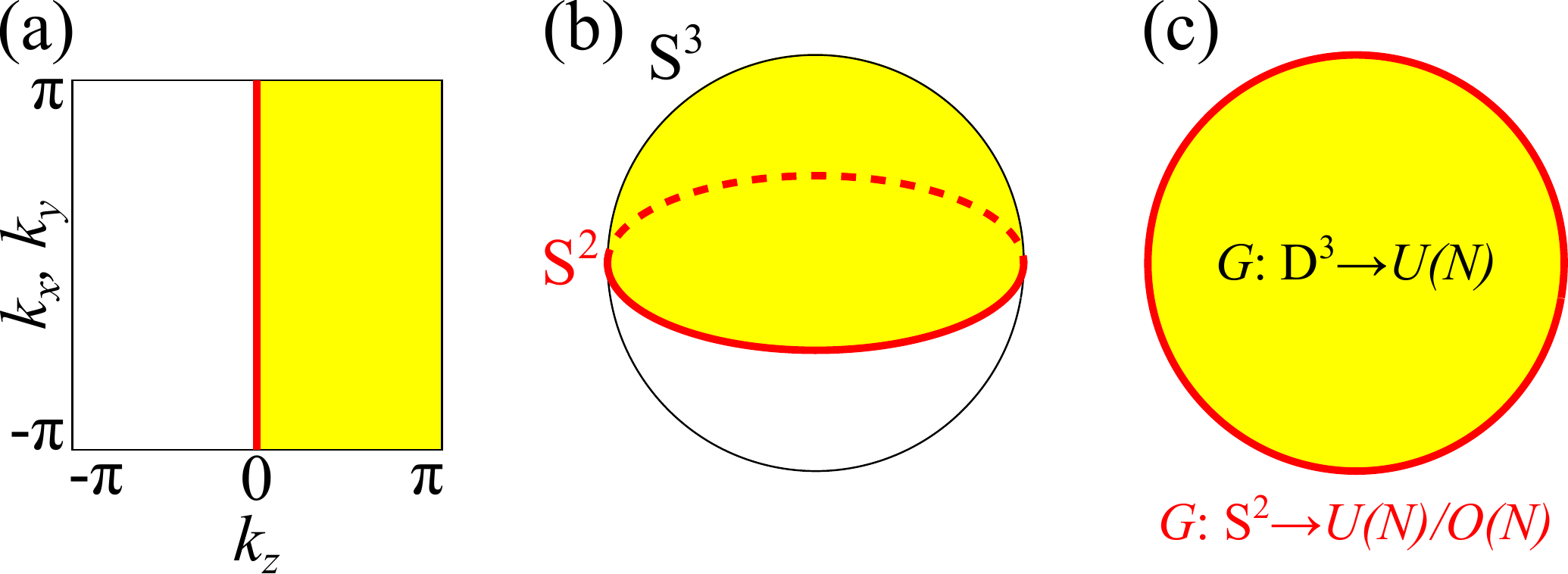}
\caption{Effective domain for the sewing matrix.
(a) A plane representing the 3D Brillouin zone.
The yellow region shows the effective Brillouin zone, and the red region with $k_z=0$ is a $C_{2z}T$-invariant plane.
The $k_z=\pi$ plane is assumed to be topologically trivial.
(b) A 3-sphere equivalent to the 3D Brillouin zone.
(c) The sewing matrix $G$ in the yellow region and on its boundary (red).
}
\label{fig.patch}
\end{figure}

Similarly, the 3D topological invariant corresponds to the third homotopy class of $G$, and it is determined by the second homotopy class of $G$ on $C_{2z}T$-invariant planes.
To show this, let us continuously deform the Brillouin torus $T^3$ as a 3-sphere $S^3$ as shown in Fig.~\ref{fig.patch}(a,b).
This is valid as long as 1D and 2D topological invariants are all trivial, because then the noncontractible 1D and 2D cycles can be shrunk to a point without changing the 3D topological invariant as long as the three-dimensional manifold itself is not shrunk to a point.
To make the 3-sphere in Fig.~\ref{fig.patch}(b) from the 3-torus Fig.~\ref{fig.patch}(a), we deform each torus at a fixed $k_z$ ($\ne \pm \pi$) to a sphere, and that at $k_z=\pm \pi$ to a point.
Then, the $k_z=0$ plane becomes the equator, and the $k_z=\pm \pi$ planes becomes the north and south poles, respectively.
Under $C_{2z}T$, the wave functions on the northern hemisphere $\simeq D^3$ are transformed to the wave functions on the southern hemisphere and vice versa, while the equator $\simeq S^2$ is invariant.
Therefore, the effective domain consists of the upper hemisphere and its boundary as shown in Fig.~\ref{fig.patch}(b).
The relevant homotopy group for the effective domain is the relative homotopy group $\pi_3[M,X]$, which classifies the maps $D^3\rightarrow M$ under the constraint $\d D^3=S^2\rightarrow X\subset M$, where $D^3$ is a three-dimensional disk and $\d D^3=S^2$ is the boundary of $D^3$~\cite{hatcher2002algebraic,turner2012quantized,trifunovic2017bott,sun2018conversion}.
In our case, $M=U(N)$, $X=U(N)/O(N)$, $D^3$ is the upper hemisphere, and $S^2$ is the equator as shown in Fig.~\ref{fig.patch}(c).
The relative homotopy group $\pi_3\left[U(N),U(N)/O(N)\right]$ has the form $\pi_3\left[U(N)\right]\times \pi_2\left[U(N)/O(N)\right]$ as shown in Appendix~\ref{sec.homotopy_sequence}.
Since the $\pi_3[U(N)]$ part comes from the gauge degrees of freedom, the third homotopy class is determined by the  second homotopy class on the invariant subspace, i.e., $\pi_2\left[U(N)/O(N)\right]\simeq \pi_1[O(N)]$,
\begin{align}
\label{eq.3-homotopy}
\frac{\pi_3\left[U(N),U(N)/O(N)\right]}{\pi_3\left[U(N)\right]}
\simeq
\pi_2\left[U(N)/O(N)\right]
\end{align}
We demonstrate this relation in Sec.~\ref{sec.third}.

%%%%%%%%%%%%%%%%%%%%%%%%%%%%%%%%%%%%%%%%   FIGURE   %%%%%%%%%%%%%%%%%%%%%%%%%%%%%%%%%%%%%%%%%%%%%%%%%%%%%%
%%%%%%%%%%%%%%%%%%%%%%%%%%%%%%%%%%%%%%%%   FIGURE   %%%%%%%%%%%%%%%%%%%%%%%%%%%%%%%%%%%%%%%%%%%%%%%%%%%%%%

\section{The first homotopy class}
\label{sec.first}

Here, we review the correspondence between the 1D winding number of $G$ in a smooth gauge and the first Stiefel-Whitney number $w_1$ in a real gauge~\cite{ahn2018band}, since the same idea is used to derive the correspondence between the second homotopy class of $G$ and the second Stiefel-Whitney number in the next section.

Let us suppose that $\ket{u_{n\bf k}}$ is smooth and the sewing matrix $G$ is defined in this basis. Then, we perform a gauge transformation to get new basis states
$\ket{\tilde{u}_{nk}}=U_{mn}(k)\ket{u_{mk}}$ such that
$\tilde{G}(k)=U^{\dagger}(k)G(k)U^*(k)$ and $U(k)$ is smooth for $0< k< 2\pi$, where $0\le k<2\pi$ parametrizes a closed loop in the $C_{2z}T$-invariant plane.
If we require the reality condition $\tilde{G}(k)=1$ for the new basis, we have $\det [U^{\dagger}(k)G(k)U^*(k)]=\det \tilde{G}(k)=1$, so $\d_k\log \det U(k)=\frac{1}{2}\d_k\log\det G(k)$.
We have a transition function $t_{mn}\equiv \braket{\tilde{u}_{m0}|\tilde{u}_{n2\pi}}=U^*_{pm}(0)U_{pn}(0+2\pi)$ since $\braket{u_{p 0}|u_{q2\pi}}=\delta_{pq}$ due to the smoothness of the original basis.
Its determinant is given by the winding number of $G$, which we write as $w$, namely,
$\det t=\det [U^*(0)U(2\pi)]=\exp [\int^{2\pi}_{0} \d_k\log \det U(k)]=\exp [\frac{1}{2}\int^{2\pi}_{0} \d_k\log \det G(k)]=(-1)^w$.
As the first Stiefel-Whitney number $w_1$ is defined by $(-1)^{w_1}=\det t$, we have $w_1=w$ modulo 2.

The above construction shows the relation Eq.~\eqref{eq.smooth=real}. Here, $\det t=\pm 1$ characterizes $\pi_0[O(N)]$ because $t\in O(N)$, and $\exp [\int^{2\pi}_{0} \d_k\log \det U(k)]=\pm 1$ characterizes the gauge-invariant part of $\pi_1[U(N)/O(N)]$.
Let us explain more about this.
Although $U$ is not periodic when $\det t=-1$ because then $\det U$ is antiperiodic, $U$ is periodic as an element of $U(N)/O(N)$ (recall $U(2\pi)=U(0)t$).
Smooth gauge transformations can change the winding number of $U$, but it does not change the periodic condition of $U$.
Therefore, among nontrivial elements in $\pi_1[U(N)/O(N)]$, only the loops along which $\det U$ changes sign is robust against gauge transformations.

\section{The second homotopy class}
\label{sec.second}

In this section, we show that the second homotopy class of $G$ in a smooth gauge corresponds to the second Stiefel-Whitney number in a real gauge.
Below we begin with the definition of the second Stiefel-Whitney number in a real gauge, and then go to a smooth gauge.
The gauge transformation matrix is associated with the sewing matrix by Eq.~\eqref{eq.sewing=U}.

%%%%%%%%%%%%%%%%%%%%%%%%%%%%%%%%%%%%%%%%   FIGURE   %%%%%%%%%%%%%%%%%%%%%%%%%%%%%%%%%%%%%%%%%%%%%%%%%%%%%%
\begin{figure}[t!]
\includegraphics[width=8.5cm]{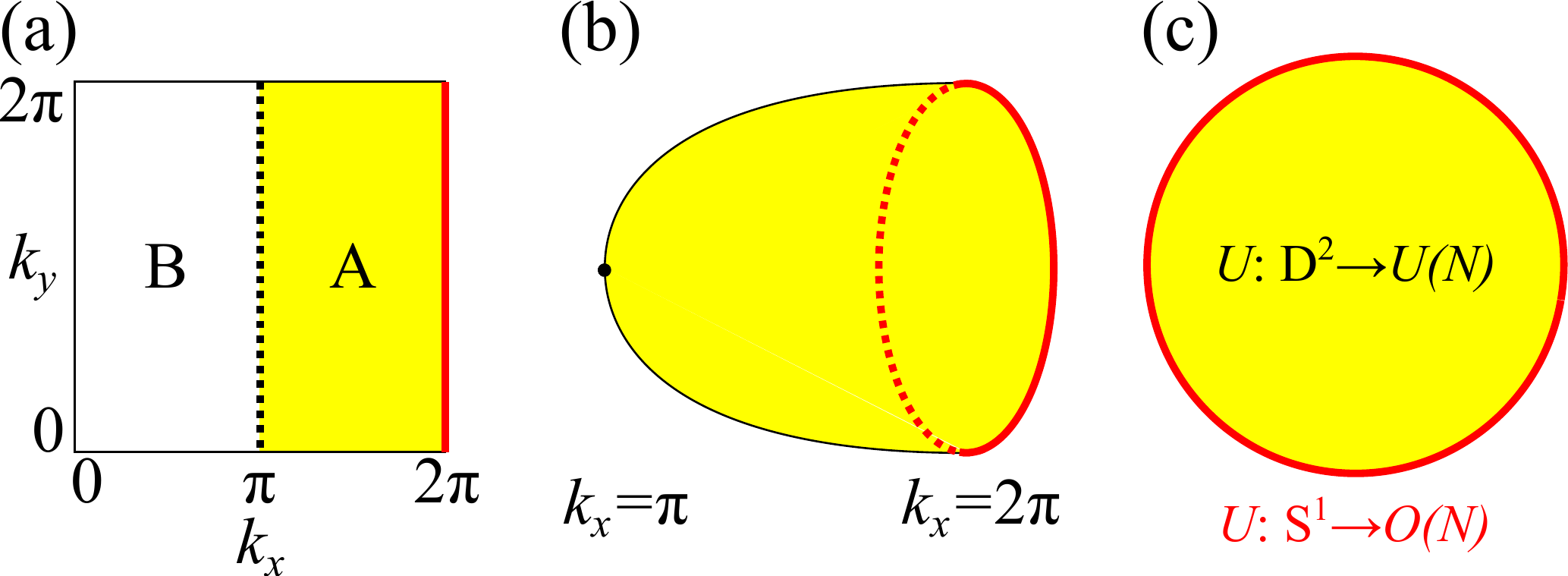}
\caption{Gauge transformation from a real to a smooth complex gauge in a $C_{2z}T$-invariant plane.
(a) $C_{2z}T$-invariant 2D Brillouin zone covered by two patches $A$ and $B$ in a real gauge.
(b) The patch $A$ whose $k_x=\pi$ line is contracted to a point.
(c) The gauge transformation matrix $U$ on the patch $A$.
}
\label{fig.patch2}
\end{figure}
%%%%%%%%%%%%%%%%%%%%%%%%%%%%%%%%%%%%%%%%   FIGURE   %%%%%%%%%%%%%%%%%%%%%%%%%%%%%%%%%%%%%%%%%%%%%%%%%%%%%%

\begin{figure}[t!]
\includegraphics[width=8.5cm]{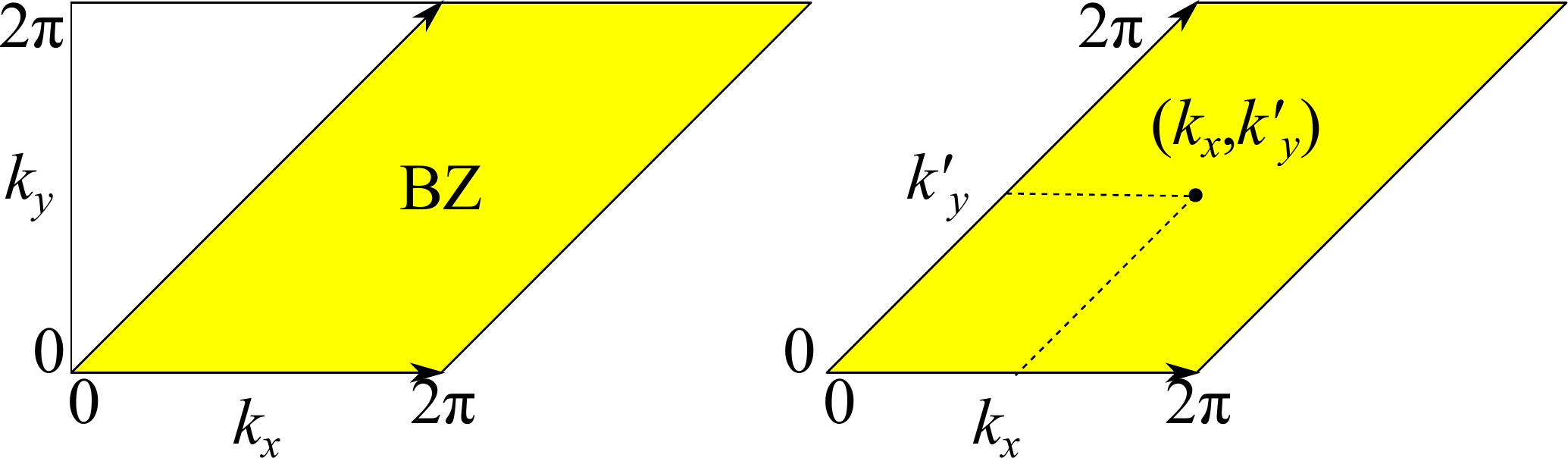}
\caption{
Dehn twist of the Brillouin zone.
A Brillouin zone defined by $0\le k_x,k_y\le 2\pi$ is Denn-twisted to a tilted Brillouin zone shown as a yellow (shaded) region.
When both 1D cycles along $k_x$ and $k_y$ have nontrivial 1D topological invariants, i.e., $w_{1x}=w_{1y}=1$, we have a trivial 1D cycle $k_y'$ by a Denn twist because $w_{1y'}=w_{1x}+w_{1y}=0 \mod 2$.}
\label{fig.Dehn-twist}
\end{figure}

We take a real gauge and cover the Brillouin zone torus with two patches $A$ and $B$, overlapping on the lines $k_x=0$ and $k_x=\pi$ [See Fig.~\ref{fig.patch2}(a)].
When the first Stiefel-Whitney numbers are nontrivial along both $k_x$ and $k_y$ directions, we should introduce more patches so that there exist discontinuous transitions along the $k_y$ direction~\cite{ahn2018band}.
However, we can always Dehn twist the Brillouin zone in those cases as shown in Fig.~\ref{fig.Dehn-twist} such that only one cycle has nontrivial $w_1$ at most, and we take the nontrivial cycle to be along the $k_x$ direction~\cite{ahn2018band}.
We assume that such a Dehn twist is done.
Also, we take the transition function at $k_x=\pi$ to be trivial.
That is, we require that real occupied states $\ket{\tilde{u}_{n\bf k}}$ are smooth within the patches, but there can exist a nontrivial transition function on the equator defined by
\begin{align}
t^{AB}_{mn}(k_y)
\equiv \braket{\tilde{u}^A_{m (2\pi,k_y)}|\tilde{u}^B_{n(0,k_y)}},
\end{align}
which is an element of the orthogonal group $O(N)$ for $N$ occupied bands.
The second Stiefel-Whitney number $w_2$ is defined by the 1D winding number of the transition function $t^{AB}$ modulo 2.

Then, we consider a gauge transformation to smooth states $\ket{u_{n(k_x,k_y)}}$ via
\begin{align}
\label{eq.real-to-smooth}
\ket{u_{n(k_x,k_y)}}
&=U_{mn}(k_x,k_y)\ket{\tilde{u}^{A}_{m(k_x,k_y)}}, \; \pi\le k_x\le 2\pi,\notag\\
\ket{u_{n(k_x,k_y)}}
&=U_{mn}(k_x,k_y)\ket{\tilde{u}^{B}_{m(k_x,k_y)}}, \; 0\le k_x\le \pi,
\end{align}
where $U(k_x,k_y)$ is smooth for $0\le k_x,k_y\le 2\pi$.
The gauge transformation matrix $U$ satisfies
\begin{align}
t^{AB}_{mn}(k_y)
&=\braket{\tilde{u}^A_{m(2\pi,k_y)}|\tilde{u}^B_{n(0,k_y)}}\notag\\
&=U_{mp}(2\pi,k_y)\braket{u_{p(2\pi,k_y)}|u_{q(0,k_y)}}U^{*}_{nq}(0,k_y)\notag\\
&=U_{mp}(2\pi,k_y)\delta_{pq}U^{*}_{nq}(0,k_y),
\end{align}
where we used that $\ket{u_{n\bf k}}$ is smooth in the last line.
By choosing a gauge $U(0,k_y)=1$,
we have
\begin{align}
\label{eq.kx=0}
U(2\pi,k_y)=t^{AB}(k_y)\in O(N).
\end{align}
We further require that $U(\pi,k_y)$ is independent of $k_y$, i.e.,
\begin{align}
\label{eq.kx=pi}
U(\pi,k_y)=U_0\in U(N),
\end{align}
as shown in Fig.~\ref{fig.patch2}.
It is possible to take this gauge because the 1D topological invariant, the first Stiefel-Whitney number, is trivial along the $k_y$ direction.

%%%%%%%%%%%%%%%%%%%%%%%%%%%%%%%%%%%%%%%%   FIGURE   %%%%%%%%%%%%%%%%%%%%%%%%%%%%%%%%%%%%%%%%%%%%%%%%%%%%%%
%%%%%%%%%%%%%%%%%%%%%%%%%%%%%%%%%%%%%%%%   FIGURE   %%%%%%%%%%%%%%%%%%%%%%%%%%%%%%%%%%%%%%%%%%%%%%%%%%%%%%

Now, the information on the wave function topology, encoded in the transition function $t^{AB}$ in a real gauge, is reflected in the unitary matrix $U$ under the constraint of Eq.~\eqref{eq.kx=0}.
Since $U$ is constant on $k_x=0$ and $k_x=\pi$ lines, the lines can be shrunk to a point as long as topology is concerned.
After the shrinking, the $B$ region becomes a sphere, and the $A$ region becomes a cap as shown in Fig.~\ref{fig.patch2}(b).
All possible $U$s are homotopically equivalent in the region $B$ because they are classified by the homotopy group $\pi_2[U(N)]=0$.
Therefore, we only need to study the homotopy class of $U$ on the region $A$.
The homotopy group of $U$ on the region $N$ with the boundary condition Eq.~\eqref{eq.kx=pi} is the relative homotopy group $\pi_2[U(N),O(N)]$ [See Fig.~\ref{fig.patch2}(c)].
Here, $[U(N),O(N)]$ means that $U\in U(N)$ inside the region $A$ and $U\in O(N)$ on its boundary, which is the equator.
Because $\pi_2[U(N)]=0$, the relative homotopy class of $U$ is in one-to-one correspondence with the homotopy class of $U$ on its boundary, which is nothing but the homotopy class of the transition function $t^{AB}\in \pi_1[O(N)]$.
That is, $\pi_2[U(N),O(N)]\simeq \pi_1[O(N)]$.
Moreover, the relative homotopy group of $U$ is isomorphic to the homotopy group of $G=UU^T$.
In other words, $\pi_2 [U(N),O(N)]\simeq \pi_2[U(N)/O(N)]$~\cite{hatcher2002algebraic}, where the isomorphism is provided by the projection from $[U(N),O(N)]$ to $U(N)/O(N)$.
Therefore,
\begin{align}
\pi_2[U(N)/O(N)]\simeq \pi_1[O(N)].
\end{align}
As the homotopy groups for smooth and periodic gauge transfomations are trivial, the process described here provides an explicit mapping for the isomorphism in Eq.~\eqref{eq.smooth=real} in the $d=2$ case.

Using this formulation of the second Stiefel-Whitney number as a homotopy class of the sewing matrix, we can simply derive the unique characteristic of the Stiefel-Whitney numbers, the Whitney sum formula~\cite{ahn2018band,hatcher2003vector}, if we require some natural algebraic rules for the second homotopy classes on the Brillouin zone torus.
Let us first consider a real gauge and suppose that the occupied bands are grouped into blocks ${\cal B}_i$ of bands isolated from each other, so that different blocks are not connected by transition functions.
For example, transition functions are block-diagonalized when there are finite energy gaps between blocks, though a gapped energy spectrum is not necessary in general to have a block-diagonal form of transition functions.
On the Brillouin zone torus having noncontractible 1D cycles along $k_x$ and $k_y$ directions, the second Stiefel-Whitney number of the whole occupied bands $\oplus {\cal B}_i$ is related to the Stiefel-Whitney numbers of blocks by the {\it Whitney sum formula}~\cite{ahn2018band,hatcher2003vector}
\begin{align}
\label{eq.Whitney}
w_2(\oplus_i{\cal B}_i)
&=\sum_iw_2({\cal B}_i)
+\sum_{i\ne j}w_{1}^x({\cal B}_i)w_{1}^y({\cal B}_j),
\end{align}
where $w_1^{a=x,y}$ is the first Stiefel-Whitney number along $k_{a=x,y}$.
The appearance of the second term in the summation is a unique characteristics of the second Stiefel-Whitney number.

From the relation between the transition function in a real gauge and the sewing matrix in a smooth gauge derived above, we can infer that the Whitney sum formula should be applicable to the blocks that decouple the sewing matrix in a smooth gauge.
For instance, let us consider two blocks ${\cal B}_1$ and ${\cal B}_2$ of occupied bands that block-diagonalize the sewing matrix as
\begin{align}
G({\bf k})
&=
\begin{pmatrix}
e^{i{\theta_1({\bf k})}}G_1^{(0)}({\bf k})&0\\
0&e^{i{\theta_2({\bf k})}}G_2^{(0)}({\bf k})
\end{pmatrix},
\end{align}
where the $U(1)$ factor $e^{i\theta_{i=1,2}}$ of each block is singled out.
Let $N_1$ and $N_2$ be the number of the bands in the blocks ${\cal B}_1$ and ${\cal B}_2$, respectively.
Then, the second homotopy class of $G:T^2\rightarrow U(N_1)/O(N_1)\times U(N_2)/O(N_2)$ is determined by the second homotopy classes of $G_1^{(0)}\in SU(N_1)/SO(N_1)$, $G_2^{(0)}\in SU(N_2)/SO(N_2)$, and $(e^{i\theta_1},e^{i\theta_2})\in U(1)\times U(1)\simeq T^2$.
The parities of the second homotopy class for $G_1^{(0)}$ and $G_2^{(0)}$ correspond to $w_2({\cal B}_1)$ and $w_2({\cal B}_2)$, respectively.
Because the generators of $\pi_2[SU(N_{i})/SO(N_{i})]$ for $i=1,2$ are mapped to the generators of $\pi_2[U(N_1+N_2)/O(N_1+N_2)]$ by the inclusion maps, we have $w_2({\cal B}_1\oplus {\cal B}_2)=w_2({\cal B}_1)+w_2({\cal B}_2)$ when the $U(1)\times U(1)$ part is neglected.
For the map $T^2\rightarrow U(1)\times U(1)$, we can define the degree of the map as a homotopy invariant
\begin{align}
&\frac{1}{(2\pi)^2}\int_{\rm BZ} d^2k \left(\d_{k_x}\theta_1\d_{k_y}\theta_2-\d_{k_x}\theta_2\d_{k_y}\theta_1\right)\notag\\
&=w_{1}^x({\cal B}_1)w_{1}^y({\cal B}_2)-w_{1}^x({\cal B}_2)w_{1}^y({\cal B}_1)\text{ mod 2},
\end{align}
where we used that $\theta_{i=1,2}({\bf k})$ is homotopically equivalent to $w_{1}^x({\cal B}_i)k_x+w_{1}^y({\cal B}_i)k_y$ because they have the same 1D winding number: $w_{1}^{j}({\cal B}_i)$ along $k_{j=x,y}$.
If we require that this homotopy invariant contributes to the two-dimensional topological invariant, that is, the second Stiefel-Whitney invariant, we obtain the Whitney sum formula $w_2({\cal B}_1\oplus {\cal B}_2)=w_2({\cal B}_1)+w_2({\cal B}_2)+w_{1}^x({\cal B}_1)w_{1}^y({\cal B}_2)-w_{1}^x({\cal B}_2)w_{1}^y({\cal B}_1)$. The generalization to the cases with many blocks is straightforward.

\section{Strong Topological Invariant in 3D}
\label{sec.strong}

Let us apply the results obtained above to the 3D topological insulator protected by $C_{2z}T$ symmetry.
For this, we first review the definition of the 3D topological invariant in a real gauge and study the stability condition for the corresponding topological phase.

Analogous to the Fu-Kane-Mele invariant, one can define a 3D strong topological invariant by using $w_2(0)$ and $w_{2}(\pi)$ defined on the $k_z=0$ and $k_z=\pi$ planes, respectively, as
\begin{align}
\Delta_{\rm 3D}\equiv w_2(\pi)-w_2(0),
\end{align}
which is identical to the $Z_{2}$-invariant proposed in~\cite{fang2015new}.
Because $w_2(\pi)=w_2(0)$ in weakly coupled layered systems, the nonzero $\Delta_{\rm 3D}$ is a 3D strong invariant~\cite{fang2015new}.
In this respect, the phase with $\Delta_{\rm 3D}=1$ can be called a {\it 3D strong Stiefel-Whitney insulator} (SWI).
Below, we show that the 3D strong SWI is a well-defined stable topological phase only when all the Chern numbers are trivial: $c_{1}^{xy}=c_{1}^{yz}=c_{1}^{zx}=0$ where $c_{1}^{ij}$ indicates the Chern number defined in the $k_ik_j$ plane.
Here a stable topological phase indicates that its 3D topological invariant remains intact against adding atomic insulators~\cite{po2018fragile}, whose 2D or 3D band topology is trivial (Though, their 1D invariants might be nontrivial because 1D invariants are related to the Wannier centers~\cite{read2017compactly,shiozaki2017topological}.)

To address the stability of the 3D strong SWI, let us consider the Whitney sum formula which can be applied to $C_{2z}T$-symmetric 2D BZ torus~\cite{ahn2018band,hatcher2003vector}.
According to Eq.~(\ref{eq.Whitney}), $w_2$ is fragile against adding bands with $w_2=0$ if they have nontrivial $w_1$~\cite{ahn2018band,ahn2019failure}, although it is a stable K-theory invariant: it is stable against adding bands with $w_2=0$ and $w_1=0$.
For instance, if a block ${\cal B}'$ of bands is added to the original block ${\cal B}$, $w_2$ changes as
\begin{align}
\delta w_{2}
&=w_{2}({\cal B}\oplus {\cal B}')-w_{2}({\cal B})\notag\\
&=w_{2}({\cal B}')+w_{1}^x({\cal B})w_{1}^y({\cal B}')+w_{1}^x({\cal B}')w_{1}^y({\cal B})
\end{align}
on both $k_z=0$ and $k_z=\pi$ planes.
Even when $w_{2}({\cal B}')=0$, $\delta w_2$ can be nonzero when $w_1^{x,y}({\cal B}')$ is nontrivial unless $w_{1}^x({\cal B})=w_{1}^y({\cal B})=0$.
The corresponding change of $\Delta_{\rm 3D}$ is given by
\begin{align}
\delta \Delta_{\rm 3D}
&=\Delta_{\rm 3D}({\cal B}\oplus {\cal B}')-\Delta_{\rm 3D}({\cal B})\notag\\
&=c_1^{xz}({\cal B})w_{1}^y({\cal B}')+c_1^{yz}({\cal B})w_{1}^x({\cal B}') (\text{ mod 2}),
\end{align}
where ${\cal B}'$ is assumed to be from an atomic insulator, such that $w_{1}^{x,y}({\cal B}')$ are the same in both $k_{z}=0$ and $k_{z}=\pi$ planes, and $\Delta_{\rm 3D}({\cal B}')=0$.
In order to define $\Delta_{\rm 3D}$ independent of adding atomic insulators, we should require that $c_{1}^{xz}({\cal B})=c_{1}^{yz}({\cal B})=0$.
Since $c_{1}^{xy}=0$ is always imposed by $C_{2z}T$ symmetry, we conclude that $\Delta_{\rm 3D}$ becomes a well-defined stable topological invariant only when all the Chern numbers vanish in the BZ.

\section{The third homotopy class}
\label{sec.third}

Let us now show that $\Delta_{\rm 3D}$ is in fact equivalent to the magnetoelectric polarizability $P_3$, defined by the effective Lagrangian ${\cal L}_{\rm top}=P_3{\bf E}\cdot {\bf B}$.
First, we assume that all the Chern numbers are trivial, i.e., $c_{1}^{xy}=c_{1}^{yz}=c_{1}^{zx}=0$, and thus $\Delta_{\rm 3D}$ is a stable topological invariant.
Under this assumption, we can take a smooth gauge over the whole Brillouin zone~\cite{brouder2007exponential,wang2010equivalent}.
In a smooth gauge,
$P_3$ takes the form of the 3D Chern-Simons invariant~\cite{qi2008topological,wang2010equivalent}
\begin{align}
P_3
&=\frac{1}{8\pi^2}\int_{\rm BZ}d^3k\epsilon^{ijk}{\rm Tr}\left[A_i\d_jA_k-\frac{2i}{3}A_iA_jA_k\right],
\end{align}
where $A_{mn}({\bf k})=\braket{u_{m\bf k}|i\nabla_{\bf k}|u_{n\bf k}}$ is the Berry connection.
Since
$A^*_i({\bf k})=
G^{-1}({\bf k})(C^{-1}_{2z})_{ij}A_j(-C_{2z}{\bf k})G({\bf k})
-G^{-1}({\bf k})i\nabla_{k_i}G({\bf k})$
in $C_{2z}T$-symmetric systems,
we have~\cite{schindler2018higher}
\begin{align}
2P_3=\frac{1}{24\pi^2}\int_{\rm BZ}d^3k\epsilon^{ijk}{\rm Tr}\left[(G^{-1}\d_iG)(G^{-1}\d_jG)(G^{-1}\d_kG)\right],
\end{align}
which is nothing but the 3D winding number of the sewing matrix $G$.
Let us note that, under the gauge transformation $\ket{u_{n\bf k}}\rightarrow \ket{u'_{n\bf k}}=U_{mn}({\bf k})\ket{u_{m\bf k}}$, $2P_3$ changes as 
\begin{align}
\delta(2P_3)=2\times \frac{1}{24\pi^2}\int_{\rm BZ}{\rm Tr}(U^{-1}dU)^3\in 2Z.
\end{align}
Therefore, $2P_3$ is a $Z_2$ topological invariant well-defined modulo two.

Equation~\eqref{eq.3-homotopy} implies that the 3D winding number of $G$ is determined by the second homotopy class of $G$ on two $C_{2z}T$-invariant planes with $k_z=0$ and $k_z=\pi$, respectively.
Let us prove this explicitly for the simplest case with two occupied bands ($N=2$) neglecting the $U(1)$ factor.
This assumption is good enough to determine the topological invariant modulo two because $\pi_3[U(N)]\simeq \pi_3[SU(2)]$ and $\pi_2[U(N)/O(N)]\simeq \pi_2[SU(2)/SO(2)]$ modulo two for all $N\ge 2$.

Let us note that $SU(2)\simeq S^3$ and $SU(2)/SO(2)\simeq S^2$, which can be obtained from the fact that a $SU(2)$ element $U=a_0+ia_1\sigma_1+ia_2\sigma_2+ia_3\sigma_3$ has four real coefficients $a_{0},\hdots,a_3$ satisfying $a_0^2+a_1^2+a_2^2+a_3^2=1$, and $SU(2)/SO(2)$ elements are the ones with $a_2=0$.
Then, the winding number of $G:T^3\rightarrow SU(2)\simeq S^3$ is determined by the degree of $G$, which is given by the number of points in $T^3$ that is mapped to a given element $u\in SU(2)$~\cite{wang2010equivalent}.
The degree does not depend on the choice of $u$~\cite{wang2010equivalent}.
If we choose $u\in SU(2)/SO(2)$, when a point with $k_z\ne 0$ or $\pi$ is mapped to $u$, the point has a partner related by $C_{2z}T$ that is mapped to the same element $u$ because $u$ satisfies Eq.~\eqref{eq.sewing_constraint} and $u^T=u$. In this case, a pair of points contributes an even number to the degree of $G$.
Accordingly, the parity of the degree is given by the sum of the degree computed on the $k_z=0$ and $k_z=\pi$ planes, i.e., the degree of the map $G:T^2\rightarrow SU(2)/SO(2)\simeq S^2$ on these planes.
In other words, $2P_3=w_2(\pi)+w_2(0)$ modulo two because the degree of the map on the planes is identical to the second Stiefel-Whitney number.
Using $w_2=-w_2$ modulo two, we eventually obtain
\begin{align}
\label{strong=P3}
\Delta_{\rm 3D}= 2P_3\text{ mod 2}.
\end{align}

\section{Bulk-boundary correspondence}
\label{sec.bulk-boundary}

Given the relation in Eq.~\eqref{strong=P3}, the bulk-boundary correspondence of the 3D strong SWI can be described by using the known topological effective action~\cite{qi2008topological} 
\begin{align}
S_{\rm top}^{(\rm bulk)}[{\cal A}]
&
= \frac{P_3}{16\pi}\int dtd^3x \epsilon^{ijkl}{\cal F}_{ij}{\cal F}_{kl},
\end{align}
where ${\cal F}_{ij}=\d_i{\cal A}_j-\d_j{\cal A}_i$ is the electromagnetic field strength, and we take $\hbar=c=e=1$.

To study the boundary effect, let us consider a geometry with a 3D strong SWI on one side with $x<0$ and the vacuum on the other side with $x>0$, which is modeled by $P_3(t,{\bf x})=P_3\Theta(-x)$.
After integrated by parts, the effective action can be written as a boundary action,
\begin{align}
S_{\rm top}[{\cal A}]
=&\frac{1}{16\pi}\int_{{\bb R}^4} dtd^3x P_3(t,{\bf x}) \epsilon^{ijkl}\d_i(4{\cal A}_j\d_k{\cal A}_l),\notag\\
=&\frac{P_3}{4\pi}\int_{x=0} dtd^2x \epsilon^{ijk}{\cal A}_i\d_j{\cal A}_k,
\end{align}
which means that the bulk topological term induces the surface quantum Hall effect with Hall conductivity $\sigma_H^{(\rm surf)}=P_3/2\pi$.
In other words, the Chern number on the surface is given by 
\begin{align}
c_1^{(\rm surf)}=P_3\text{ mod 1},
\end{align}
because $\sigma_H=c_1/2\pi$.
The surface state with $c_1^{(\rm surf)}=1/2$ can be realized in two different ways depending on the symmetry of the system.
Namely, it can be either a Chern insulator with half-quantized Hall conductance as in axion insulators~\cite{essin2009magnetoelectric} or a semimetal with an odd number of Dirac points as in time-reversal-invariant 3D topological insulators~\cite{fu2007topological}.

Here we consider the case where $C_{2z}$ and $T$ symmetries are broken individually whereas the combined symmetry $C_{2z}T$ is preserved.
If both $C_{2z}$ and $T$ are the symmetries of the system, the 3D strong SWI is not allowed when $T^2=1$~\cite{fang2015new} because $w_{2}(0)=w_{2}(\pi)$ as $C_{2z}$ eigenvalues indicate~\cite{ahn2018band}.
On the other hand, when $T^{2}=-1$, $\Delta_{3D}=1$ is allowed, but it is identical to the well known Fu-Kane-Mele invariant~\cite{fang2015new} since the bulk $P_3$ is nontrivial~\cite{wang2010equivalent}.

In a 3D strong SWI, both gapless and gapped states appear on the surface as shown in Fig.~\ref{fig.scheme}(b).
To understand this, let us consider an orthorhombic geometry.
On the top and bottom surfaces, which are $C_{2z}T$-invariant, insulating states are not allowed because $C_{2z}T$ symmetry requires the vanishing of the Chern number.
Instead, there appears an odd number of Dirac points, whose $\pi$ Berry phase is protected due to the quantization of the Berry phase by $C_{2z}T$~\cite{fang2015new}.
On the other hand, side surfaces is gapped because $C_{2z}T$ symmetry is broken and thus 2D Dirac points cannot be protected. So the side surfaces become Chern insulators with half-quantized Hall conductance with $c_1=n\pm 1/2$ where $n$ is an integer.
The sign of $c_1$ on the side surfaces are related by $C_{2z}T$ symmetry through
\begin{align}
c_1({\bf x})=-c_1(C_{2z}{\bf x}),
\end{align}
where $c_1=(1/2\pi)\int_{BZ}d^2k{\rm Tr}{\bf F}\cdot \hat{\bf n}$, and $\hat{\bf n}$ is the surface normal unit vector pointing outwards, and ${\bf F}=d{\bf A}-i{\bf A}\times {\bf A}$ is the Berry curvature.
It means that the front side surface and the back side surface form two domains with different Chern numbers.
Therefore, chiral 1D states appear on side hinges that are the boundaries of the two different domains.
Let us note that the stability condition for the 3D strong SWI, that is, the vanishing of the bulk Chern numbers, prohibits other anomalous surface states on the side surfaces, and thus the chiral hinge states can become well-localized. 

%%%%%%%%%%%%%%%%%%%%%%%%%%%%%%%%%%%%%%%%   FIGURE   %%%%%%%%%%%%%%%%%%%%%%%%%%%%%%%%%%%%%%%%%%%%%%%%%%%%%%
\begin{figure}[t!]
\includegraphics[width=8.5cm]{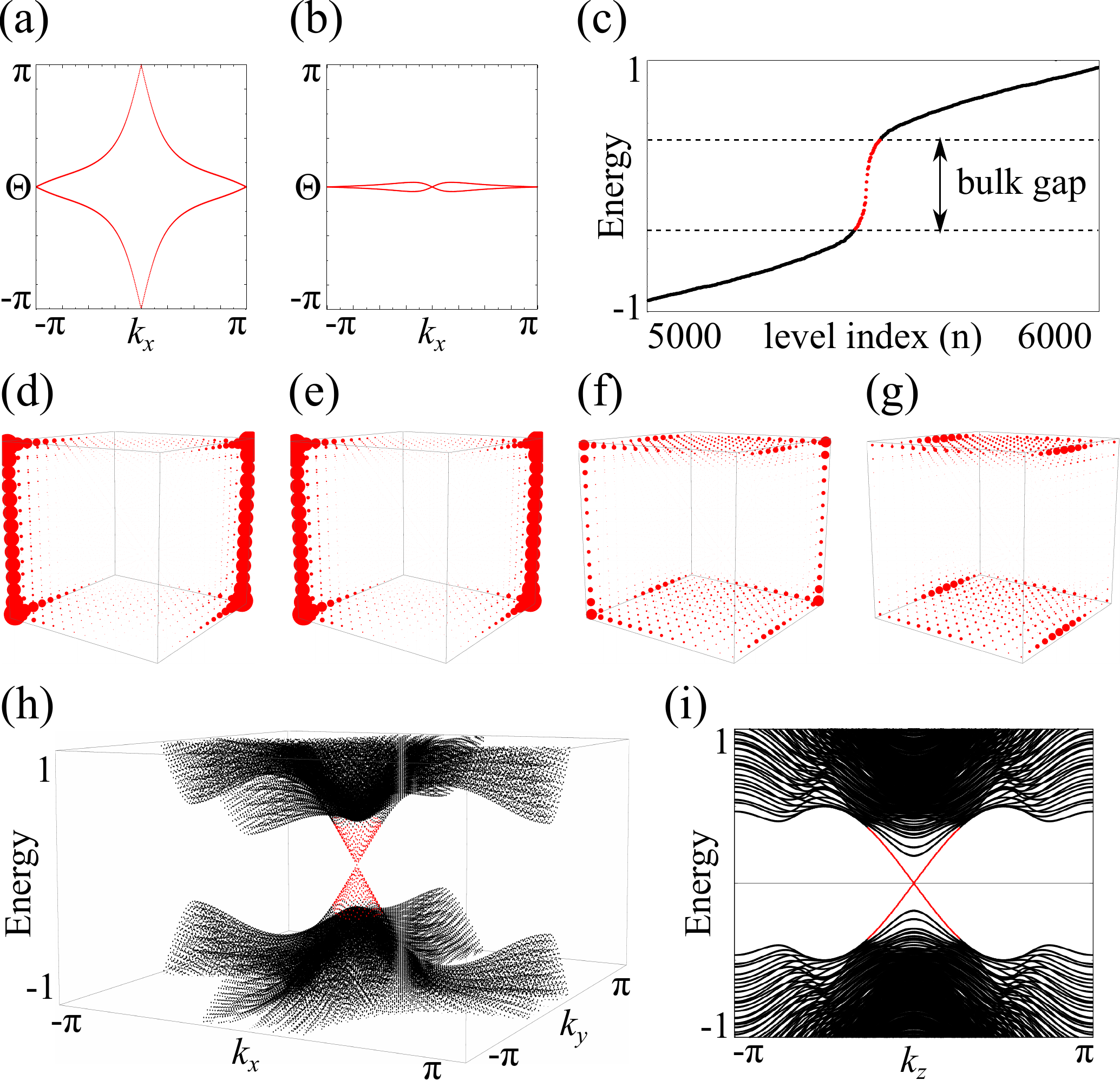}
\caption{Numerical calculation using the model in Eq.~\eqref{eq.TB}.
$v=0.5$, $m_{14}=0.5$, and $m_{24}=0.3$.
(a,b) Wilson loop calculation on (a) $k_z=0$ and (b) $k_z=\pi$ planes.
(c-g) Finite-size calculation with $14\times 14\times 14$ unit cells.
(c) Energy eigenvalues in an increasing order.
Near the half filling ($n=2\times 14^3=5488$), anomalous states appear within the bulk gap $|E_g|\approx 0.37$.
(d-g) Density profile of in-gap states computed by using the
(d) first
(e) second
(f) thid
(g) fourth
highest occupied state below Fermi level at half filling.
(h) Band structure of a model which is periodic in the $(x,y)$-direction, and has 20 unit cells in the $z$-direction.
The Dirac point in the bulk band gap originates from the top and bottom surfaces
(i) Band structure of the model which has $20\times 20$ unit cells in the $(x,y)$-direction but periodic in the $z$-direction.
Hinge states appear within the bulk band gap.
}
\label{fig.TB}
\end{figure}
%%%%%%%%%%%%%%%%%%%%%%%%%%%%%%%%%%%%%%%%   FIGURE   %%%%%%%%%%%%%%%%%%%%%%%%%%%%%%%%%%%%%%%%%%%%%%%%%%%%%%

\section{Tight-binding model}
\label{sec.TB}

To illustrate the higher-order band topology of a 3D strong SWI, let us perform a tight-binding model analysis. We start with a Hamiltonian describing a 3D Dirac semimetal with both $PT$ and $C_{2z}T$ symmetries, where $P$ indicates spatial inversion,
\begin{align}
H_{\rm DSM}
=
&\sin k_x\Gamma_1+\sin k_y \Gamma_2\notag\\
&+(-2+\cos k_x+\cos k_y+\cos k_z)\Gamma_3,
\end{align}
where $\Gamma_1=\sigma_x$, $\Gamma_2=\tau_y\sigma_y$, and $\Gamma_3=\sigma_z$. The symmetry representations for $PT$ and $C_{2z}T$ are given by $PT=K$ and $C_{2z}T=K$.
Two Dirac points appear at $(k_x,k_y,k_z)=(0,0,\pm \pi/2)$, respectively, each of which is protected by $PT$ symmetry.
Each Dirac point carries a $Z_2$ monopole charge~\cite{morimoto2014weyl,fang2015topological,zhao2017pt}, which is identical to the nontrivial $w_2$ on a closed manifold wrapping the Dirac point. Since the energy gap is finite except at $(k_x,k_y,k_z)=(0,0,\pm \pi/2)$, the sphere wrapping a Dirac point can be deformed to two parallel planes with $k_z=0$ and $k_{z}=\pi$, respectively. Then the monopole charge of a Dirac point is given by the difference of $w_2(0)$ and $w_2(\pi)$.

Adding $C_{2z}T$-preserving perturbations that open the bulk and surface band gaps, we have
\begin{align}
\label{eq.TB}
H
=H_{\rm DSM}+v \sin k_z\Gamma_4+m_{14}\Gamma_{14}+m_{24}\Gamma_{24},
\end{align}
where $\Gamma_{4}=\tau_x\sigma_y$, $\Gamma_{14}=\tau_x\sigma_z$, and $\Gamma_{24}=-\tau_z$.
$v\ne 0$ opens the bulk gap because it breaks $PT$ symmetry whereas
$PT$-preserving terms $m_{14}\ne 0$ and $m_{24}\ne 0$ open the gap on the side surfaces (a closely related model was constructed in~\cite{geier2018second}).
Since $m_{14}$ and $m_{24}$ deform each Dirac point to a monopole nodal line in the bulk Brillouin zone, $H$ can be regarded as a monopole nodal line semimetal with an additional $PT$-breaking parameter $v$ that transforms as the in-plane magnetic field.
As long as the perturbations are small such that the band gap does not close on $k_z=0$ and $k_z=\pi$ planes, $w_2(\pi)-w_2(0)=1$ mod 2 should be maintained.
Wilson loop calculations in Fig.~\ref{fig.TB}(a,b) show that $w_2(0)=1$ and $w_2(\pi)=0$ because $w_2$ is given by the number of linear crossing points at $\Theta=\pi$ modulo two, where $\Theta$ is the phase of the eigenvalue of the Wilson loop operator~\cite{yu2011equivalent,fang2015topological,zhao2017pt,bzdusek2017robust,
ahn2018band,song2018all,bouhon2018wilson,bradlyn2018disconnected} [See also Appendix~\ref{sec.Wilson}].
Our finite-size calculations in Fig.~\ref{fig.TB}(c) with $14\times 14\times 14$ unit cells show that the system has anomalous in-gap states.
The hinge and surface states coexist as shown in Fig.~\ref{fig.TB}(d-g), which are from the first four highest energy occupied states below the Fermi level at half filling.
The linearly dispersing spectrum of the in-gap states is visible if we calculate the band structure with partial open boundary conditions as shown in Fig.~\ref{fig.TB}(h,i).

\section{Discussion}
\label{sec.discussion}

We have shown that the second Stiefel-Whitney number is a homotopy invariant that determines the second homotopy class of the sewing matrix for $C_{2z}T$.
Taking into account the related results for $d=1$ and $3$ reported before, we conclude that the $d$-dimensional topological invariant is the measure of the $d$th homotopy class of the sewing matrix in $C_{2z}T$-symmetric systems.

Since the homotopy equivalence of the matrix representation of symmetry groups classifies topological crystalline insulators in principle~\cite{
bradlyn2017topological}
%,cano2018building,bradlyn2018band,
%elcoro2017double,vergniory2017graph,vergniory2018high
, we expect that all known topological crystalline insulators can be expressed as a homotopy invariant of some sewing matrix.
For instance, the Fu-Kane invariant~\cite{fu2006time} is indeed defined as the homotopy invariant of the sewing matrix.
The mirror Chern number can also be interpreted as the second homotopy class of the sewing matrix [see Appendix~\ref{sec.mirrorChern}].
Another example is the magnetoelectric polarizability.
In general, the magnetoelectric polarizability is quantized in the presence of a symmetry operation that reverses the space-time orientation regardless of whether it is symmorphic or nonsymmorphic~\cite{varjas2015bulk}.
When the magnetoelectric polarizability is quantized by a symmorphic symmetry, it is known to be expressed by the 3D winding number of the sewing matrix~\cite{schindler2018higher}.
We further show in Appendix~\ref{sec.quantization_P3} that the same is true for nonsymmorphic symmetries.
Namely, the topological invariants protected by a nonsymmorphic symmetry can also be described as the obstruction to a momentum-independent representation of the relevant sewing matrix.
Extending the analysis to include other magnetic space groups is definitely one interesting direction for future research.

\begin{acknowledgments}
{\it Acknowledgments.|}
We thank Benjamin J. Wieder, Bogdan Andrei Bernevig, and Yoonseok Hwang for helpful comments to our manuscript.
J.A. was supported by IBS-R009-D1.
B.-J.Y. was supported by the Institute for Basic Science in Korea (Grant No. IBS-R009-D1) and Basic Science Research Program through the National Research Foundation of Korea (NRF) (Grant No. No.0426-20190008), and  the POSCO Science Fellowship of POSCO TJ Park Foundation (No.0426-20180002).
This work was supported in part by the US Army Research Office under Grant Number W911NF-18-1-0137.

{\it Note added.|}
During the preparation of our manuscript, we have found a related manuscript~\cite{benjamin2018the} that also identifies the presence of chiral hinge states in $C_{2z}T$-protected insulators with $P_3=1$.
\end{acknowledgments}

\appendix

\section{Some properties of homotopy groups}
\label{sec.homotopy_sequence}

In this Appendix, we prove some properties of homotopy groups we use in the main text.
The main tool to be used is the long exact sequence of homotopy groups~\cite{hatcher2002algebraic,turner2012quantized,sun2018conversion}:
\begin{align}
\label{eq.les1}
...
&\xrightarrow{\d_{p+1}} \pi_{p}(X)
\xrightarrow{i_p^*} \pi_p(M)
\xrightarrow{j_p^*} \pi_p(M,X)\notag\\
&\xrightarrow{\d_p} \pi_{p-1}(X)
\xrightarrow{i_p^*}
...,
\end{align}
where $i_p:X\rightarrow M$ and $j_p:M\rightarrow (M,X)$ are inclusions, $i^*_p$ and $j^*_p$are maps for homotopy groups induced by $i_p$ and $j_p$, and $\d$ is the restriction to the boundary.
This sequence is {\it exact} because the image of a map is the kernel of the next map, e.g., ${\rm im}\,i_p^*=\ker j_p^*$.
It is also valid when $\pi_p(M,X)$ is substituted by $\pi_p(M/X)$ because the two homotopy groups are isomorphic~\cite{hatcher2002algebraic,sun2018conversion}.
\begin{align}
\label{eq.les2}
\hdots
&\xrightarrow{\d_{p+1}} \pi_{p}(X)
\xrightarrow{i_p^*} \pi_p(M)
\xrightarrow{j_p^*} \pi_p(M/X)\notag\\
&\xrightarrow{\d_p} \pi_{p-1}(X)
\xrightarrow{i_p^*}
\hdots.
\end{align}

\subsection{Equivalence between real and smooth gauges}

Let us prove Eq.~\eqref{eq.smooth=real}, that is, $\pi_d[U(N)/O(N)]/{\rm im\,}j^*_d \simeq \pi_{d-1}[O(N)]$ when $d\ne 4n$ for a positive integer $n$.
It can be proved for arbitrary $N$ when $d=1,2$, which are dimensions studied in the main text, whereas we need the large $N$ limit in general dimensions,
This follows from the exact sequence in Eq.~\eqref{eq.les2}.
In our case, $M=U(N)$, and $X=O(N)$.
We have
\begin{align}
\hdots\rightarrow &\pi_d[U(N)]
\xrightarrow{j_d^*} \pi_d[U(N)/O(N)]\notag\\
&\xrightarrow{\d_{d}} \pi_{d-1}[O(N)]
\xrightarrow{i_{d-1}^*}
\pi_{d-1}[U(N)]\rightarrow \hdots.
\end{align}
Then, we have
\begin{align}
\frac{\pi_d[U(N)/O(N)]}{{\rm im\,} j^*_d}
\simeq
\ker i^*_{d-1},
\end{align} 
where we used the exactness of maps $\ker \d_d={\rm im}\, j_d^*$ and ${\rm im}\,\d_d=\ker i_{d-1}^*$ and the group isomorphism theorem $\pi_d[U(N)/O(N)]/\ker \d_d\simeq {\rm im\,}\d_d$.
Notice that $i_{d-1}^*$ is a trivial map for $d\ne 4n$ for a positive integer $n$ when $N$ is large enough.
When $d$ is odd, it is because $\pi_{d-1}[U(N)]=0$ for $d\le 2N$.
In particular, $\pi_0[U(N)]=\pi_2[U(N)]=0$ for all $N$.
When $d=2$, $i_{d-1}^*$ is trivial because orthogonal group elements have quantized determinants, $+1$ or $-1$, so that they cannot have a winding of the determinant (recall that $\pi_1[U(N)]$ is characterized by the winding number of the determinant of the unitary matrix).
When $d=6$, the map $i_{5}^*$ is trivial because $\pi_{5}[O(N)]=0$.
Bott periodicity then shows that the same is true for $2+8m$ and $6+8m$ dimensions for a positive integer $m$ when $N$ is large enough, i.e., $d\le 2N$ and $d\le N-1$. 
On the other hand, when $d=4n$ for a positive integer $n$, $i_{d-1}^*$ is not trivial. 
This is related to the fact that the reality condition on wave functions (equivalently, $PT$ symmetry) does not require that the $2n$-th Chern class vanishes, and the Chern class in a real gauge is called the Pontrjagin class~\cite{nakahara2003geometry}.
Let us recall that the $2n$-th Chern number is given by the $(4n-1)$-th nontrivial homotopy of the transition function.
The $2n$-th Chern class of real wave functions does not vanish because the nontrivial homotopy class of the transition function in $\pi_{d-1}[O(N)]$ survives as an element in $\pi_{d-1}[U(N)]$.
Accordingly, 
\begin{align}
\ker i^*_{d-1}\simeq \pi_{d-1}[O(N)]\text{ for }d\notin 4{\bb Z}_+,
\end{align}
where ${\bb Z}_+$ is the set of positive integers.
This finishes the proof.

\subsection{Third homotopy group of $C_{2z}T$}

Let us prove that
\begin{align}
\pi_3\left[U(N),\frac{U(N)}{O(N)}\right]=\pi_3\left[U(N)\right]\times \pi_2\left[\frac{U(N)}{O(N)}\right].
\end{align}
This follows from the long exact sequence
\begin{align}
...
&\xrightarrow{\d_{4}^*} \pi_3\left[\frac{U(N)}{O(N)}\right]
\xrightarrow{i_3^*} \pi_3\left[U(N)\right]
\xrightarrow{j_3^*} \pi_3\left[U(N),\frac{U(N)}{O(N)}\right]\notag\\
&\xrightarrow{\d_3}\pi_2\left[\frac{U(N)}{O(N)}\right]
\xrightarrow{i_2^*} \pi_2\left[U(N)\right]
\xrightarrow{j_2^*} ...
\end{align}
and that the map ${\rm Im}\,i^*_n=0$ for $n=2,3$.
${\rm Im}\,i^*_2=0$ because $\pi_2\left[U(N)\right]=0$, and ${\rm Im}\,i^*_3=0$ because ${\rm Tr}(U^{-1}dU)^3=0$ for $U\in U(N)/O(N)$.

\section{Wilson loop method}
\label{sec.Wilson}

%%%%%%%%%%%%%%%%%%%%%%%%%%%%%%%%%%%%%%%%   FIGURE   %%%%%%%%%%%%%%%%%%%%%%%%%%%%%%%%%%%%%%%%%%%%%%%%%%%%%%
\begin{figure}[t!]
\includegraphics[width=8.5cm]{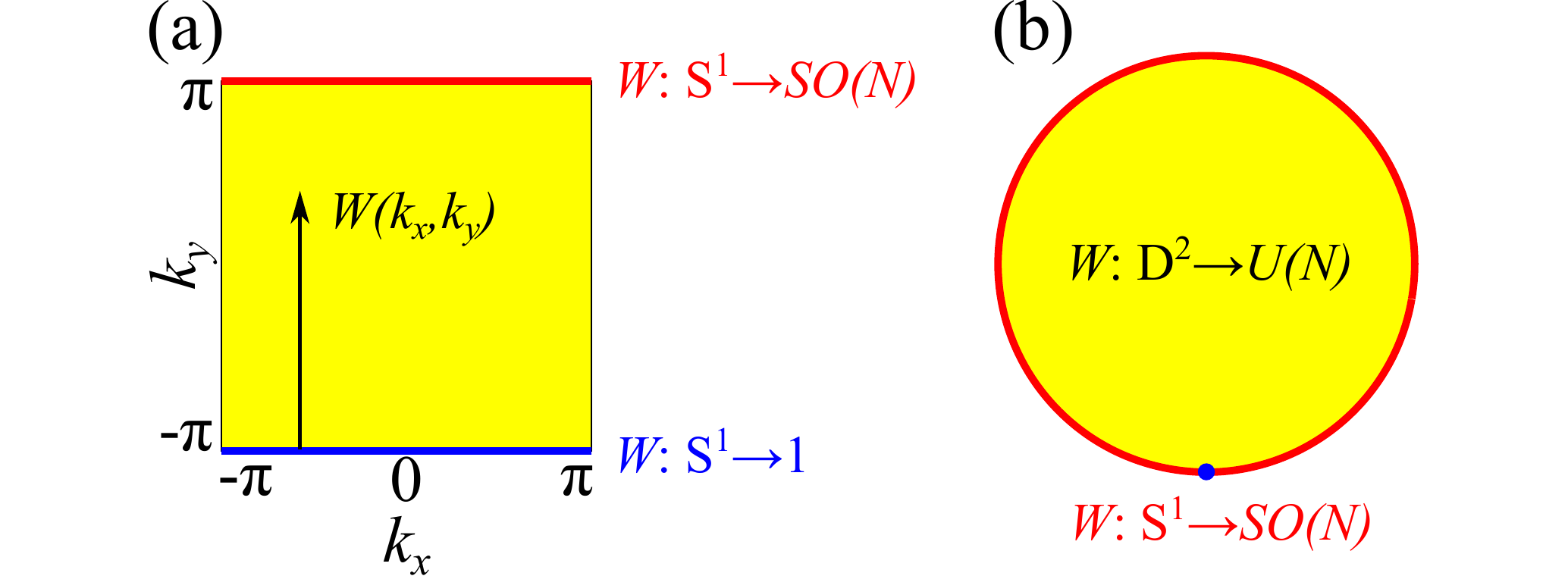}
\caption{Wilson line operator in a $C_{2z}T$-invariant Brillouin zone.
(a) $W(k_x,k_y)$ is the Wilson line operator $W_{(k_x,-\pi)\rightarrow (k_x,k_y)}$.
(b) Deformation of (a) after $k_x=-\pi$, $k_x=\pi$, and $k_y=-\pi$ lines are contracted to a point.
}
\label{fig.Wilson}
\end{figure}
%%%%%%%%%%%%%%%%%%%%%%%%%%%%%%%%%%%%%%%%   FIGURE   %%%%%%%%%%%%%%%%%%%%%%%%%%%%%%%%%%%%%%%%%%%%%%%%%%%%%%

In this Appendix, we show the connection between the second homotopy class of the sewing matrix for $C_{2z}T$ and the winding number of the Wilson loop spectrum in an invariant plane.
This provides a new insight into the Wilson loop method~\cite{yu2011equivalent,fang2015topological,zhao2017pt,bzdusek2017robust,
ahn2018band,song2018all,bouhon2018wilson,bradlyn2018disconnected}.

We first define a Wilson line operator for the occupied states on the line connecting ${\bf k}$ and ${\bf k}'$ by
\begin{align}
W_{{\bf k}\rightarrow {\bf k}'}
&=\lim_{{\bm \delta}\rightarrow 0}
F_{{\bf k}'-{\bm \delta}}F_{{\bf k}'-2{\bm \delta}}...F_{{\bf k}+{\bm \delta}}F_{\bf k},
\end{align}
where
\begin{align}
(F_{\bf k})_{mn}
=\braket{u_{m{\bf k}+{\bm \delta}}|u_{n\bf k}},
\end{align}
and $m,n$ are indices for occupied states.
The transition matrix $F$ satisfies the following equation in $C_{2z}T$-symmetric systems.
\begin{align}
(F^*_{\bf k})_{mn}
&=\braket{u_{m{\bf k}+{\bm \delta}}|u_{n\bf k}}^*\notag\\
&=\braket{C_{2z}Tu_{m{\bf k}+{\bm \delta}}|C_{2z}Tu_{n\bf k}}\notag\\
&=G^*_{pm}({\bf k}+{\bm \delta})\braket{u_{p{\bf k}+{\bm \delta}}|u_{q\bf k}}G_{qn}({\bf k})\notag\\
&=[G^{\dagger}({\bf k}+{\bm \delta})F_{\bf k}G({\bf k})]_{mn}.
\end{align}
It follows that
\begin{align}
W_{{\bf k}\rightarrow {\bf k}'}^*
&=\lim_{{\bm \delta}\rightarrow 0}
F^*_{{\bf k}'-{\bm \delta}}F^*_{{\bf k}'-2{\bm \delta}}...F^*_{{\bf k}+{\bm \delta}}F^*_{\bf k}\notag\\
&=G^{\dagger}({\bf k}')
W_{{\bf k}\rightarrow {\bf k}'}G({\bf k}).
\end{align}
Therefore, we find that
\begin{align}
G({\bf k}')
=W_{{\bf k}\rightarrow {\bf k}'}G({\bf k})W_{{\bf k}\rightarrow {\bf k}'}^T
\end{align}

For simplicity, we assume that all 1D topological invariants are trivial.
Then, we can take a gauge $G(k_x,-\pi)=1$ such that
\begin{align}
G(k_x,k_y)
=W_{(k_x,-\pi)\rightarrow (k_x,k_y)}W_{(k_x,-\pi)\rightarrow (k_x,k_y)}^T
\end{align}
Because we are in a smooth gauge, we have
\begin{align}
1
&=G(k_x,-\pi)\notag\\
&=G(k_x,\pi)\notag\\
&=W_{(k_x,-\pi)\rightarrow (k_x,\pi)}W_{(k_x,-\pi)\rightarrow (k_x,\pi)}^T,
\end{align}
so the Wilson loop operator belong to the orthogonal group at $k_y=\pi$:
\begin{align}
W_{(k_x,-\pi)\rightarrow (k_x,\pi)}\in SO(N).
\end{align}
It belongs to $SO(N)\subset O(N)$ because it is continously connected to the identity element $W_{(k_x,-\pi)\rightarrow (k_x,-\pi)}=1$.

Let us contract the $k_x=-\pi$, $k_x=\pi$, and $k_y=-\pi$ lines to a point, which is possible due to the assumption that the 1D topology is trivial, as shown in Fig.~\ref{fig.Wilson}(a,b).
As we show in Appendix.~\ref{sec.homotopy_sequence}, the relative homotopy class of $\pi_2[U(N),SO(N)]$ of $W(k_x,k_y)\equiv W_{(k_x,-\pi)\rightarrow (k_x,k_y)}$ is determined by its homotopy class on the boundary $\pi_1[SO(N)]$.
Notice that the relative homotopy class of $W$ is one-to-one correspondence with the second homotopy class of $G$ as derived in Appendix.~\ref{sec.homotopy_sequence}.
Also, the homotopy class in $\pi_1[SO(N)]$ is given by the winding number of the Wilson loop operator $W[k_x]\equiv W_{(k_x,-\pi)\rightarrow (k_x,\pi)}$.
Therefore, we conclude that the second homotopy class of $G(k_x,k_y)$ is in one-to-one correspondence with the 1D winding number of the Wilson loop operator $W[k_x]$.
In practice, one obtains the winding number of the Wilson loop operator from the winding pattern of its spectrum, which can be calculated in a gauge-invariant way.

\section{Mirror Chern number}
\label{sec.mirrorChern}

Here, we show that the mirror Chern number can be represented as the second homotopy class of the sewing matrix for the mirror operator $M$.
We follow the same procedure we used in the $C_{2z}T$-symmetric case.

Consider the sewing matrix for the $z$-mirror operator $M$ in a two-dimensional system, where ${\bf k}=(k_x,k_y)$.
\begin{align}
G_{mn}({\bf k})
=\braket{\psi_{nM\bf k}|M|\psi_{n\bf k}}
\end{align}
As mirror operator satisfies $M^2=1$ (when $M^2=-1$, we can consider $M'=iM$ that satisfies $M'^2=1$),
\begin{align}
G^{\dagger}({\bf k})=G(M{\bf k}).
\end{align}
The sewing matrix transforms by
\begin{align}
G_{mn}({\bf k})
\rightarrow G'_{mn}({\bf k})
=[U^{\dagger}(M {\bf k})G({\bf k})U({\bf k})]_{mn},
\end{align}
under the gauge transformation$
\ket{u_{n\bf k}}\rightarrow \ket{u'_{n\bf k}}
=U_{mn}({\bf k})\ket{u_{m\bf k}},
$
where $G'_{mn}({\bf k})=\braket{u'_{mM\bf k}|M|u'_{n\bf k}}$.

In a mirror-invariant plane where $M{\bf k}={\bf k}$, $G$ is a unitary Hermitian matrix, so it has the following form
\begin{align}
\label{eq.mirror_sewing_canonical}
G({\bf k})
=
U^{\dagger}({\bf k})
\begin{pmatrix}
1_{N\times N}&0\\
0&-1_{M\times M}
\end{pmatrix}
U({\bf k}),
\end{align}
where $U({\bf k})\in U(N+M)$ is a gauge transformation needed to diagonalize the sewing matrix $G({\bf k})$.
Since diagonal $U\in U(N)\times U(M)$ does not change the matrix $G$, the sewing matrix belong to the quotient space called the complex Grassmannian manifold
\begin{align}
G({\bf k})\in U(N+M)/U(N)\times U(M).
\end{align}
It can have a second homotopy class because $\pi_n[U(N+M)/U(N)\times U(M)]=0$ for $n=$odd and $Z$ for $n=$even.

Following the same logic used for $C_{2z}T$ symmetry, we can derive the relation between the mirror Chern number and the winding number of $G$.
Let us consider the same spherical geometry used for the $C_{2z}T$-symmetric case [See Fig.~\ref{fig.patch2}(a)], and start from the mirror eigenstate basis $\ket{\tilde{u}_{n(k_x,k_y)}}$ with $\tilde{G}={\rm diag}(1_{N\times N},-1_{M\times M})$.
In the eigenstate basis, the Brillouin zone is covered with two patches, $A$ and $B$, overlapping on $k_x=\pi$ and $k_x=2\pi$.
We take a gauge where the transition function is trivial on $k_x=\pi$.
Then, the transition matrix $t^{AB}(\phi)=\braket{\tilde{u}^A_{n(2\pi,k_y)}|\tilde{u}^B_{n(0,k_y)}}$ takes the form
\begin{align}
t^{AB}=
\begin{pmatrix}
t^{AB}_+&0\\
0&t^{AB}_-
\end{pmatrix}
,\quad t^{AB}_+\in U(N), t^{AB}_-\in U(M).
\end{align}
The winding number of $t^{AB}_+$ and $t^{AB}_-$ gives the Chern number of the sector with mirror eigenvalue $+1$ and $-1$, respectively. 

After we transform to a smooth gauge, the winding number of the transition function is encoded in the second homotopy class of the sewing matrix $G$.
Here, we assume that the total Chern number vanishes in order to take a smooth gauge at the cost of giving up a uniform sewing matrix.
We first show that the relative homotopy class of the gauge transformation matrix $U$ needed to go to a smooth gauge corresponds to the homotopy class of the transition function.
Then, we get the desired result by Eq.~\eqref{eq.mirror_sewing_canonical}.

Let us first show that $U$ satisfies the constraint that it equals to the transition matrix on the equator.
We consider a gauge transformation from mirror eigenstates $\ket{\tilde{u}_{n(k_x,k_y)}}$ to smooth states $\ket{u_{n(k_x,k_y)}}$ defined by
$
\ket{\tilde{u}^{A/B}_{n(k_x,k_y)}}
=U_{mn}(k_x,k_y)\ket{u_{m(k_x,k_y)}}.
$
The gauge transformation matrix $U$ satisfies
$
t^{AB}_{mn}(k_y)
=U^{*}_{pm}(2\pi,k_y)U_{pn}(0,k_y),
$
where we used that $\ket{u_{n\bf k}}$ is smooth such that $\braket{u_{p(k_x,k_y)}|u_{q(k_x,k_y)}}=\delta_{pq}$.
By choosing the gauge $U(0,k_y)=1$, we have
\begin{align}
U(2\pi,k_y)=t^{AB}(k_y)\in U(N)\times U(M),
\end{align}
on the equator.

Next, we show that the relative homotopy class of $U$ is given by the homotopy class of the transition function $t^{AB}$.
This follows from the exact sequence in Eq.~\eqref{eq.les1}.
In our case, $M=U(N+M)$, and $X=U(N)\times U(M)$.
As $\pi_2[U(N+M)]=0$, we have
\begin{align}
&0
\xrightarrow{j_2^*} \pi_2[U(N+M),U(N)\times U(M)]\notag\\
&\xrightarrow{\d_2} \pi_{1}[U(N)\times U(M)]
\xrightarrow{i_2^*}
\pi_{1}[U(N+M)]
\xrightarrow{j_1^*}...,
\end{align}
where $0=\{1\}$.
Then, $\pi_2[U(N+M),U(N)\times U(M)] \simeq \ker i_2^*$ because $\ker \d_2={\rm im}\, j_2^*=1$ and ${\rm im}\,\d_2=\ker i_2^*$.
Notice that $\ker i_2^*$ is composed of elements whose total winding number vanishes, i.e., the total Chern number is trivial, such that the nontrivial element in the group characterizes the mirror Chern number.
We can also show that the homotopy group for $G$ is in one-to-one correspondence with the relative homotopy group of $U$ in the same way as we did for $C_{2z}T$ symmetry.
In conclusion, the second homotopy class of $G$ corresponds to the mirror Chern number.

\section{Quantization of magnetoelectric polarizability}
\label{sec.quantization_P3}

In a system symmetric under the space-time-orientation-reversing transformation $g$, regardless of whether it is symmorphic or nonsymmorphic, the magnetoelectric polarizability is quantized~\cite{varjas2015bulk}.
Here we show that the magnetoelectric polarizability is given by the winding number of the sewing matrix of $g$.

Let us consider a system that is symmetric under a space-time transformation $g: ({\bf r},t)\rightarrow (O{\bf r}+{\bf t},s_gt)$, where $O$ is a point group element. 
Then, the symmetry operator $\hat{U}_g$ acts on the position operator $\hat{\bf r}$ and the pure imaginary number $i$ as
\begin{align}
\hat{U}_g^{-1}\hat{\bf r}\hat{U}_g
&=O\hat{\bf \bf r}+{\bf t},\notag\\
\hat{U}_g^{-1}i \hat{U}_g
&=s_g i.
\end{align}
The symmetry representation for the Bloch states is given by
\begin{align}
G_{mn}({\bf k})=\braket{\psi_{m s_gO\bf k}|\hat{U}_g|\psi_{n\bf k}}.
\end{align}
Accordingly, the cell-periodic part transforms by
\begin{align}
\braket{u_{m s_gO\bf k}|\hat{U}_g|u_{n\bf k}}
&=\braket{\psi_{m s_gO\bf k}|e^{i{s_gO\bf k}\cdot \hat{\bf r}}\hat{U}_ge^{-i{\bf k}\cdot \hat{\bf r}}|\psi_{n\bf k}}\notag\\
&=\braket{\psi_{m s_gO\bf k}|e^{is_gO{\bf k}\cdot \hat{\bf r}}\hat{U}_ge^{-i{\bf k}\cdot \hat{\bf r}}\hat{U}_g^{-1}\hat{U}_g|\psi_{n\bf k}}\notag\\
&=\braket{\psi_{m s_gO\bf k}|e^{is_gO{\bf k}\cdot \hat{\bf r}}e^{-is_g{\bf k}\cdot O^{-1}(\hat{\bf r}-{\bf t})}\hat{U}_g|\psi_{n\bf k}}\notag\\
&=e^{is_gO{\bf k\cdot t}}\braket{\psi_{m s_gO\bf k}|\hat{U}_g|\psi_{n\bf k}}\notag\\
&=e^{is_gO{\bf k\cdot t}}G_{mn}({\bf k})
\end{align}
such that
\begin{align}
\ket{u_{n\bf k}}
=e^{iO{\bf k}\cdot {\bf t}}\overline{G}^{s_g}_{mn}({\bf k})\hat{U}_g^{-1}\ket{u_{m s_gO\bf k}},
\end{align}
where we used the notation introduced in~\cite{alexandradinata2018no}: $\overline{f}^{s_g}=f$ for $s_g=1$ and $\overline{f}^{s_g}=f^*$ for $s_g=-1$.
Using this, one can show that the Berry connection satisfies the following symmetry constraint.
\begin{widetext}
\begin{align}
A_{mn}({\bf k})
&=\braket{u_{m\bf k}|i\nabla_{\bf k}|u_{n\bf k}}\notag\\
&=\braket{\hat{U}_g^{-1}u_{p, s_gO\bf k}|e^{-iO{\bf k}\cdot {\bf t}}\overline{G}^{-s_g}_{pm}({\bf k})i\nabla_{\bf k}e^{iO{\bf k}\cdot {\bf t}}\overline{G}^{s_g}_{qn}({\bf k})|\hat{U}_g^{-1}u_{q,s_gO\bf k}}\notag\\
&=\overline{G}^{-s_g}_{pm}({\bf k})\braket{\hat{U}_g^{-1}u_{p,s_gO\bf k}|i\nabla_{\bf k}|\hat{U}_g^{-1}u_{q,s_gO\bf k}}\overline{G}^{s_g}_{qn}({\bf k})
-\delta_{mn}O^{-1}{{\bf t}}+\overline{G}^{-s_g}_{pm}({\bf k})i\nabla_{\bf k}\overline{G}^{s_g}_{pn}({\bf k})\notag\\
&=\overline{{s_g}G^*_{pm}({\bf k})\braket{u_{p,s_gO\bf k}|i\nabla_{\bf k}|u_{q,s_gO\bf k}}G_{qn}({\bf k})
+s_gG^*_{pm}({\bf k})i\nabla_{\bf k}G_{pn}({\bf k})
-\delta_{mn}O^{-1}{{\bf t}}}^{s_g}\notag\\
&=\overline{G^*_{pm}({\bf k})O^{-1}\cdot\braket{u_{p,s_gO\bf k}|i\nabla_{s_gO\bf k}|u_{q,s_gO\bf k}}G_{qn}({\bf k})
+s_gG^*_{pm}({\bf k})i\nabla_{\bf k}G_{pn}({\bf k})
-\delta_{mn}O^{-1}{{\bf t}}}^{s_g}\notag\\
&=\overline{[s_g\left(G^{-1}({\bf k})s_gO^{-1}\cdot {\bf A}(s_gO{\bf k})G({\bf k})
+G^{-1}({\bf k})i\nabla_{\bf k}G({\bf k})\right)
-O^{-1}{{\bf t}}]_{mn}}^{s_g}\notag\\
&\equiv \overline{[s_g\left(G^{-1}({\bf k})\tilde{\bf A}({\bf k})G({\bf k})
+G^{-1}({\bf k})i\nabla_{\bf k}G({\bf k})\right)
-O^{-1}{{\bf t}}]_{mn}}^{s_g}\notag\\
&\equiv \overline{[s_g\tilde{\bf A}^G({\bf k})
-O^{-1}{{\bf t}}]_{mn}}^{s_g},
\end{align}
where we introduced two notations $\tilde{\bf A}({\bf k})=s_gO^{-1}\cdot {\bf A}(s_gO{\bf k})$ and ${\bf A}^G({\bf k})=G^{-1}({\bf k}){\bf A}({\bf k})G({\bf k})+G^{-1}({\bf k})\nabla_{\bf k}G({\bf k})$.
The magnetoelectric polarizability $P_3$ then satisfies
\begin{align}
P_3
&=\overline{P_3}^{s_g}\notag\\
&=\frac{1}{8\pi^2}\int_{\rm BZ}d^3k\epsilon^{ijk}{\rm Tr}\left[\overline{A}^{s_g}_i\d_j\overline{A}^{s_g}_k+\frac{2s_gi}{3}\overline{A}^{s_g}_i\overline{A}^{s_g}_j\overline{A}^{s_g}_k\right]\notag\\
&=\frac{1}{8\pi^2}\int_{\rm BZ}d^3k\epsilon^{ijk}{\rm Tr}\left[(s_g\tilde{\bf A}^G-O^{-1}{{\bf t}})_i\d_j(s_g\tilde{\bf A}^G-O^{-1}{{\bf t}})_k+\frac{2s_gi}{3}(s_g\tilde{\bf A}^G-O^{-1}{{\bf t}})_i(s_g\tilde{\bf A}^G-O^{-1}{{\bf t}})_j(s_g\tilde{\bf A}^G-O^{-1}{{\bf t}})_k\right]\notag\\
&=\frac{1}{8\pi^2}\int_{\rm BZ}d^3k\epsilon^{ijk}{\rm Tr}\left[\tilde{A}^G_i\d_j\tilde{A}^G_k-\frac{2i}{3}\tilde{A}^G_i\tilde{A}^G_j\tilde{A}^G_k\right]
-\frac{1}{8\pi^2}\int_{\rm BZ}d^3k\epsilon^{ijk}s_g(O^{-1}{\bf t})_i{\rm Tr}\left[\d_j\tilde{A}^G_k\right]\notag\\
&=\frac{1}{8\pi^2}\int_{\rm BZ}d^3k\epsilon^{ijk}{\rm Tr}\left[\tilde{A}_i\d_j\tilde{A}_k-\frac{2i}{3}\tilde{A}_i\tilde{A}_j\tilde{A}_k\right]
+\frac{1}{24\pi^2}\int_{\rm BZ}d^3k\epsilon^{ijk}{\rm Tr}\left[(G^{-1}\d_iG)(G^{-1}\d_jG)(G^{-1}\d_kG)\right]\notag\\
&=\frac{1}{8\pi^2}\int_{\rm BZ}d^3k\epsilon^{ijk}(s_gO^{-1})_{ia}(s_gO^{-1})_{jb}(s_gO^{-1})_{kc}
{\rm Tr}\left[A_a\d_bA_c(s_gO{\bf k})-\frac{2i}{3}A_aA_bA_c(s_gO{\bf k})\right]
+\frac{1}{24\pi^2}\int {\rm Tr}(G^{-1}dG)^3\notag\\
&=\frac{1}{8\pi^2}\int_{\rm BZ}d^3k\epsilon^{abc}\det(s_gO^{-1})
{\rm Tr}\left[A_a\d_bA_c(s_gO{\bf k})-\frac{2i}{3}A_aA_bA_c(s_gO{\bf k})\right]
+\frac{1}{24\pi^2}\int {\rm Tr}(G^{-1}dG)^3\notag\\
&=\frac{1}{8\pi^2}\int_{\rm BZ}d^3k\epsilon^{abc}s_g\det O^{-1}
{\rm Tr}\left[A_a\d_bA_c-\frac{2i}{3}A_aA_bA_c\right]
+\frac{1}{24\pi^2}\int {\rm Tr}(G^{-1}dG)^3\notag\\
&=\frac{s_g\det O^{-1}}{8\pi^2}\int_{\rm BZ}d^3k\epsilon^{abc}{\rm Tr}\left[A_a\d_bA_c-\frac{2i}{3}A_aA_bA_c\right]
+\frac{1}{24\pi^2}\int {\rm Tr}(G^{-1}dG)^3\notag\\
&=(s_g\det O^{-1})P_3
+\frac{1}{24\pi^2}\int_{\rm BZ}d^3k\epsilon^{ijk}{\rm Tr}\left[(G^{-1}\d_iG)(G^{-1}\d_jG)(G^{-1}\d_kG)\right],
\end{align}
\end{widetext}
where we assumed that all the first Chern numbers are trivial to remove the term $\int d^3k \epsilon^{ijk}(O^{-1}{\bf t})_i{\rm Tr}F_{jk}$ and the total derivative term $\int d^3k \epsilon^{ijk}{\rm Tr}[\d_i(G^{-1}\d_jGA_k)]$~\cite{nakahara2003geometry} in the fifth line.
We have obtained that
\begin{align}
2P_3=\frac{1}{24\pi^2}\int_{\rm BZ}{\rm Tr}(G^{-1}dG)^3\in Z.
\end{align}
for the symmetry operation with $s_g\det O^{-1}=-1$.

\section{Anomalous boundary states of axion insulators}

%%%%%%%%%%%%%%%%%%%%%%%%%%%%%%%%%%%%%%%%   FIGURE   %%%%%%%%%%%%%%%%%%%%%%%%%%%%%%%%%%%%%%%%%%%%%%%%%%%%%%
\begin{figure}[t!]
\includegraphics[width=8.5cm]{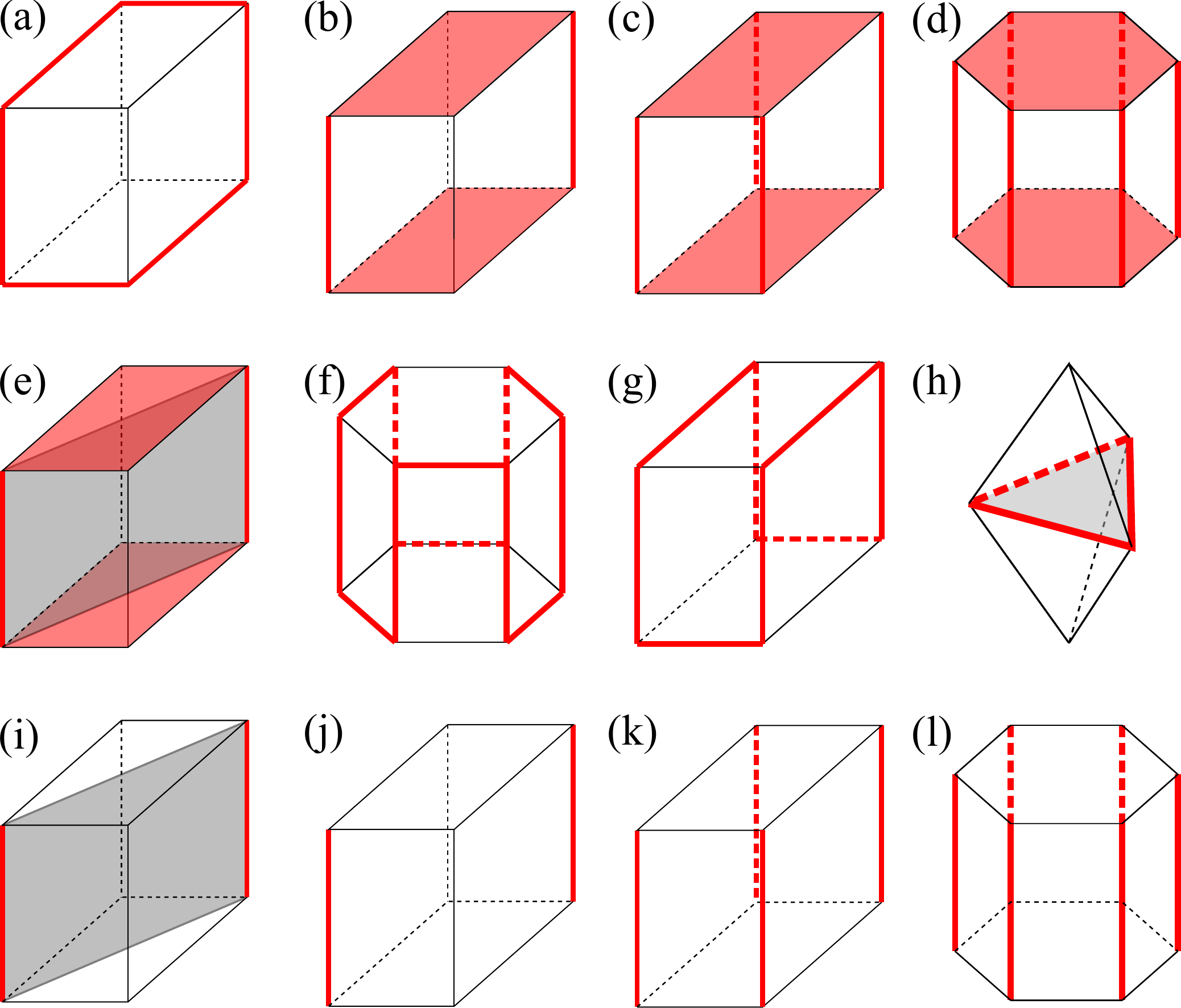}
\caption{Anomalous boundary states of axion insulators.
(a-g) Symmorphic symmetries.
(a) $P$.
(b) $C_{2z}T$.
(c) $C_{4z}T$.
(d) $C_{6z}T$.
(e) $M=C_2P$.
(f) $C_3P$.
(g) $C_4P$.
(h) $C_6P$.
(i-l) Nonsymmorphic symmetries.
(i) glide $M$.
(j) screw $C_{2z}T$.
(k) screw $C_{4z}T$.
(l) screw $C_{6z}T$.
Shaded grey regions are the (glide) mirror-invariant planes.
}
\label{fig.surface}
\end{figure}
%%%%%%%%%%%%%%%%%%%%%%%%%%%%%%%%%%%%%%%%   FIGURE   %%%%%%%%%%%%%%%%%%%%%%%%%%%%%%%%%%%%%%%%%%%%%%%%%%%%%%

Let us comment on the general bulk-boundary correspondence of insulators with quantized magnetoelectric polarizability, the so-called axion insulators.
As shown in Appendix.~\ref{sec.quantization_P3}, the magnetoelectric polarizability is quantized by a space-time-orientation-reversing symmetry in general.
Let $g$ be orientation-reversing.
Then, on the surface,
\begin{align}
c_1({\bf x})=-c_1(g{\bf x}),
\end{align}
because $c_1=(1/2\pi)\int_{BZ}d^2k{\rm Tr}{\bf F}\cdot \hat{\bf n}$, and ${\bf F}\cdot \hat{\bf n}$ changes sign by operations that reverses the space-time orientation.
Here, $\hat{\bf n}$ is the surface normal unit vector pointing outwards, and ${\bf F}=d{\bf A}-i{\bf A}\times {\bf A}$ is the Berry curvature.
Accordingly, ${\bf x}$ and $g{\bf x}$ belongs to different surface domains with opposite signs of Chern numbers if they are gapped.
Using this, we can generate the real space configuration of boundary states of axion insulators protected by space-time-orientation-reversing symmetries.
Figure~\ref{fig.surface} shows anomalous boundary states of axion insulators~\cite{fang2017rotation,khalaf2018higher,kooi2018inversion,
vanmiert2018higher,
yue2019symmetry,schindler2018higher,
ezawa2018strong,ezawa2018magnetic}.

\section{Surface states of Dirac semimetals}

In this Appendix, we identify the bulk-symmetry-preserving terms that can gap out the surface states of a $PT$-symmetric spinless Dirac semimetal.

First, consider the effective Hamiltonian for a Dirac semimetal with two Dirac points at $(0,0,k_z)=(0,0,\pm k_*)$:
\begin{align}
H_{k_z}(k_x,k_y)=k_x\Gamma_1+k_y\Gamma_2+(k^2_*-k^2)\Gamma_3,
\end{align}
where $P=\sigma_x$, $T=\tau_xK$, and
\begin{align}
\Gamma_1=\sigma_y,\quad
\Gamma_2=\tau_z\sigma_z,\quad
\Gamma_3=\sigma_x.
\end{align}
At a fixed $k_z$, the Hamiltonian describes a 2D Stiefel-Whitney insulator (normal insulator) when $m_0>0$ $(m_0<0)$ if we define $m_0=k^2_*-k_z^2$.

We now investigate the edge states of the Stiefel-Whitney insulator by considering a system occupying only a half space $x>0$, following Ref.~\cite{murakami2007tuning}.
As the $x$-direction is not periodic, we write the Hamiltonian in real space for the direction.
\begin{align}
H
&=(-i\d_x)\Gamma_1+k_y\Gamma_2+m(x)\Gamma_3\notag\\
&=
\begin{pmatrix}
k_y&m-\d_x&0&0\\
m+\d_x&-k_y&0&0\\
0&0&-k_y&m-\d_x\\
0&0&m+\d_x&k_y
\end{pmatrix}
\end{align}
where $m(x\gg0)=m_0$, $m(x\ll 0)=-m_0$, $m(x)$ changes sign at $x=0$, and $m_0>0$.

As the Hamiltonian is block-diagonal, we first solve the Schr$\ddot{\rm o}$dinger equation $Hu=Eu$ for the upper block using
$u=(u_1,u_2,0,0)^T$:
\begin{align}
\label{edge_eq1}
(m-\d_x)u_2
&=(E-k_y)u_1,\notag\\
(m+\d_x)u_1
&=(E+k_y)u_2.
\end{align}
Applying $u_2^*(m-\d_x)$ to the upper and $u_1^*(m+\d_x)$ to the lower equation, we get
\begin{align}
\label{edge_eq2}
|(m+\d_x)u_1|^2
&=(E^2-k^2_y)|u_1|^2,\notag\\
|(m-\d_x)u_2|^2
&=(E^2-k^2_y)|u_2|^2,
\end{align}
using the anti-Hermiticity of $\d_x$ which follows from the Hermiticity of $H$.
From this we find $|E|\ge |k_y|$.

We seek solutions satisfying the lowest bound $E=\pm k_y$ because we are interested in the in-gap states that are the closest to the Fermi level.
For $E=\pm k_y$, Eq.~(\ref{edge_eq1}) and (\ref{edge_eq2}) becomes
\begin{align}
(E-k_y)u_1=(m-\d_x)u_2
&=0,\notag\\
(E+k_y)u_2=(m+\d_x)u_1
&=0.
\end{align}
We have $u_2=0$ and $(m+\d_x)u_1=0$ when $E=k_y$, and $u_1=0$ and $(m-\d_x)u_2=0$ when $E=-k_y$.
Therefore, the edge state, which exponentially decays into the bulk, is
\begin{align}
u
\propto \exp\left(-\int^x ds\; m(s)\right)(1,0,0,0)^T,
\end{align}
and its energy eigenvalue is $E=k_y$.
We can do the same for the lower block to have
\begin{align}
u
\propto \exp\left(-\int^x ds\; m(s)\right)(0,0,1,0)^T,
\end{align}
and its energy eigenvalue is $E=-k_y$.
Thus, we have two edge states of opposite chirality:
\begin{align}
H^{\rm edge}_{k_z}=
\begin{pmatrix}
k_y&0\\
0&-k_y
\end{pmatrix}
\end{align}
As they exist for every $k_z$ such that $|k_z|<k^*$, these edge states form a double Fermi arcs on the surface $x=0$ of the Dirac semimetal.

Next, we include other terms as perturbations.
Define
\begin{align}
\Gamma_4=\tau_x\sigma_z,\quad
\Gamma_5=\tau_y\sigma_z,\quad
\Gamma_{ij}=\frac{[\Gamma_{i},\Gamma_j]}{2i}.
\end{align}
Then $\Gamma_{ij}$ terms with $i=1,2,3$ and $j=4,5$ are $PT$-symmetric.
Projecting the terms to the edge states
\begin{align}
u_+&= (1,0,0,0)^T,\notag\\
u_-&= (0,0,0,1)^T,
\end{align}
we find
\begin{align}
\Gamma_{14}^{\rm edge}
&=0,\notag\\
\Gamma_{15}^{\rm edge}
&=0,\notag\\
\Gamma_{24}^{\rm edge}
&=
\begin{pmatrix}
0&-i\\
i&0
\end{pmatrix}
,\notag\\
\Gamma_{25}^{\rm edge}
&=
\begin{pmatrix}
0&1\\
1&0
\end{pmatrix}
,\notag\\
\Gamma_{34}^{\rm edge}
&=0,\notag\\
\Gamma_{35}^{\rm edge}
&=0.\notag\\
\end{align}
$\Gamma_{24}$ and $\Gamma_{25}$ serve as mass terms of $H^{\rm edge}$.

Similarly, we find that $\Gamma_{14}$ and $\Gamma_{15}$ serve as mass terms if we take the $y$ direction finite.
In conclusion, all the surfaces normal to $\hat{x}$ or $\hat{y}$ are gapped if we include, e.g., $\Gamma_{14}$ and $\Gamma_{24}$ mass terms as is done in the main text.

%\bibliography{StrongSWI_ref}

\begin{thebibliography}{65}%
\makeatletter
\providecommand \@ifxundefined [1]{%
 \@ifx{#1\undefined}
}%
\providecommand \@ifnum [1]{%
 \ifnum #1\expandafter \@firstoftwo
 \else \expandafter \@secondoftwo
 \fi
}%
\providecommand \@ifx [1]{%
 \ifx #1\expandafter \@firstoftwo
 \else \expandafter \@secondoftwo
 \fi
}%
\providecommand \natexlab [1]{#1}%
\providecommand \enquote  [1]{``#1''}%
\providecommand \bibnamefont  [1]{#1}%
\providecommand \bibfnamefont [1]{#1}%
\providecommand \citenamefont [1]{#1}%
\providecommand \href@noop [0]{\@secondoftwo}%
\providecommand \href [0]{\begingroup \@sanitize@url \@href}%
\providecommand \@href[1]{\@@startlink{#1}\@@href}%
\providecommand \@@href[1]{\endgroup#1\@@endlink}%
\providecommand \@sanitize@url [0]{\catcode `\\12\catcode `\$12\catcode
  `\&12\catcode `\#12\catcode `\^12\catcode `\_12\catcode `\%12\relax}%
\providecommand \@@startlink[1]{}%
\providecommand \@@endlink[0]{}%
\providecommand \url  [0]{\begingroup\@sanitize@url \@url }%
\providecommand \@url [1]{\endgroup\@href {#1}{\urlprefix }}%
\providecommand \urlprefix  [0]{URL }%
\providecommand \Eprint [0]{\href }%
\providecommand \doibase [0]{http://dx.doi.org/}%
\providecommand \selectlanguage [0]{\@gobble}%
\providecommand \bibinfo  [0]{\@secondoftwo}%
\providecommand \bibfield  [0]{\@secondoftwo}%
\providecommand \translation [1]{[#1]}%
\providecommand \BibitemOpen [0]{}%
\providecommand \bibitemStop [0]{}%
\providecommand \bibitemNoStop [0]{.\EOS\space}%
\providecommand \EOS [0]{\spacefactor3000\relax}%
\providecommand \BibitemShut  [1]{\csname bibitem#1\endcsname}%
\let\auto@bib@innerbib\@empty
%</preamble>
\bibitem [{\citenamefont {Hasan}\ and\ \citenamefont
  {Kane}(2010)}]{hasan2010colloquium}%
  \BibitemOpen
  \bibfield  {author} {\bibinfo {author} {\bibfnamefont {M.~Z.}\ \bibnamefont
  {Hasan}}\ and\ \bibinfo {author} {\bibfnamefont {C.~L.}\ \bibnamefont
  {Kane}},\ }\bibfield  {title} {\enquote {\bibinfo {title} {Colloquium:
  topological insulators},}\ }\href@noop {} {\bibfield  {journal} {\bibinfo
  {journal} {Reviews of Modern Physics}\ }\textbf {\bibinfo {volume} {82}},\
  \bibinfo {pages} {3045} (\bibinfo {year} {2010})}\BibitemShut {NoStop}%
\bibitem [{\citenamefont {Qi}\ and\ \citenamefont
  {Zhang}(2011)}]{qi2011topological}%
  \BibitemOpen
  \bibfield  {author} {\bibinfo {author} {\bibfnamefont {X.-L.}\ \bibnamefont
  {Qi}}\ and\ \bibinfo {author} {\bibfnamefont {S.-C.}\ \bibnamefont {Zhang}},\
  }\bibfield  {title} {\enquote {\bibinfo {title} {Topological insulators and
  superconductors},}\ }\href@noop {} {\bibfield  {journal} {\bibinfo  {journal}
  {Reviews of Modern Physics}\ }\textbf {\bibinfo {volume} {83}},\ \bibinfo
  {pages} {1057} (\bibinfo {year} {2011})}\BibitemShut {NoStop}%
\bibitem [{\citenamefont {Chiu}\ \emph {et~al.}(2016)\citenamefont {Chiu},
  \citenamefont {Teo}, \citenamefont {Schnyder},\ and\ \citenamefont
  {Ryu}}]{chiu2016classification}%
  \BibitemOpen
  \bibfield  {author} {\bibinfo {author} {\bibfnamefont {C.-K.}\ \bibnamefont
  {Chiu}}, \bibinfo {author} {\bibfnamefont {J.~C.~Y.}\ \bibnamefont {Teo}},
  \bibinfo {author} {\bibfnamefont {A.~P.}\ \bibnamefont {Schnyder}}, \ and\
  \bibinfo {author} {\bibfnamefont {S.}~\bibnamefont {Ryu}},\ }\bibfield
  {title} {\enquote {\bibinfo {title} {Classification of topological quantum
  matter with symmetries},}\ }\href@noop {} {\bibfield  {journal} {\bibinfo
  {journal} {Reviews of Modern Physics}\ }\textbf {\bibinfo {volume} {88}},\
  \bibinfo {pages} {035005} (\bibinfo {year} {2016})}\BibitemShut {NoStop}%
\bibitem [{\citenamefont {Ando}\ and\ \citenamefont
  {Fu}(2015)}]{ando2015topological}%
  \BibitemOpen
  \bibfield  {author} {\bibinfo {author} {\bibfnamefont {Y.}~\bibnamefont
  {Ando}}\ and\ \bibinfo {author} {\bibfnamefont {L.}~\bibnamefont {Fu}},\
  }\bibfield  {title} {\enquote {\bibinfo {title} {Topological crystalline
  insulators and topological superconductors: from concepts to materials},}\
  }\href@noop {} {\bibfield  {journal} {\bibinfo  {journal} {Annu. Rev.
  Condens. Matter Phys.}\ }\textbf {\bibinfo {volume} {6}},\ \bibinfo {pages}
  {361--381} (\bibinfo {year} {2015})}\BibitemShut {NoStop}%
\bibitem [{\citenamefont {Shiozaki}\ \emph {et~al.}(2017)\citenamefont
  {Shiozaki}, \citenamefont {Sato},\ and\ \citenamefont
  {Gomi}}]{shiozaki2017topological}%
  \BibitemOpen
  \bibfield  {author} {\bibinfo {author} {\bibfnamefont {K.}~\bibnamefont
  {Shiozaki}}, \bibinfo {author} {\bibfnamefont {M.}~\bibnamefont {Sato}}, \
  and\ \bibinfo {author} {\bibfnamefont {K.}~\bibnamefont {Gomi}},\ }\bibfield
  {title} {\enquote {\bibinfo {title} {Topological crystalline materials:
  General formulation, module structure, and wallpaper groups},}\ }\href@noop
  {} {\bibfield  {journal} {\bibinfo  {journal} {Phys. Rev. B}\ }\textbf
  {\bibinfo {volume} {95}},\ \bibinfo {pages} {235425} (\bibinfo {year}
  {2017})}\BibitemShut {NoStop}%
\bibitem [{\citenamefont {Ahn}\ \emph {et~al.}(2018{\natexlab{a}})\citenamefont
  {Ahn}, \citenamefont {Kim}, \citenamefont {Kim},\ and\ \citenamefont
  {Yang}}]{ahn2018band}%
  \BibitemOpen
  \bibfield  {author} {\bibinfo {author} {\bibfnamefont {J.}~\bibnamefont
  {Ahn}}, \bibinfo {author} {\bibfnamefont {D.}~\bibnamefont {Kim}}, \bibinfo
  {author} {\bibfnamefont {Y.}~\bibnamefont {Kim}}, \ and\ \bibinfo {author}
  {\bibfnamefont {B.-J.}\ \bibnamefont {Yang}},\ }\bibfield  {title} {\enquote
  {\bibinfo {title} {Band topology and linking structure of nodal line
  semimetals with Z2 monopole charges},}\ }\href@noop {} {\bibfield  {journal}
  {\bibinfo  {journal} {Phys. Rev. Lett.}\ }\textbf {\bibinfo {volume} {121}},\
  \bibinfo {pages} {106403} (\bibinfo {year} {2018}{\natexlab{a}})}\BibitemShut
  {NoStop}%
\bibitem [{\citenamefont {Fang}\ and\ \citenamefont {Fu}(2015)}]{fang2015new}%
  \BibitemOpen
  \bibfield  {author} {\bibinfo {author} {\bibfnamefont {C.}~\bibnamefont
  {Fang}}\ and\ \bibinfo {author} {\bibfnamefont {L.}~\bibnamefont {Fu}},\
  }\bibfield  {title} {\enquote {\bibinfo {title} {New classes of
  three-dimensional topological crystalline insulators: Nonsymmorphic and
  magnetic},}\ }\href@noop {} {\bibfield  {journal} {\bibinfo  {journal} {Phys.
  Rev. B}\ }\textbf {\bibinfo {volume} {91}},\ \bibinfo {pages} {161105(R)}
  (\bibinfo {year} {2015})}\BibitemShut {NoStop}%
\bibitem [{\citenamefont {Morimoto}\ and\ \citenamefont
  {Furusaki}(2014)}]{morimoto2014weyl}%
  \BibitemOpen
  \bibfield  {author} {\bibinfo {author} {\bibfnamefont {T.}~\bibnamefont
  {Morimoto}}\ and\ \bibinfo {author} {\bibfnamefont {A.}~\bibnamefont
  {Furusaki}},\ }\bibfield  {title} {\enquote {\bibinfo {title} {Weyl and dirac
  semimetals with {$Z_2$} topological charge},}\ }\href@noop {} {\bibfield
  {journal} {\bibinfo  {journal} {Phys. Rev. B}\ }\textbf {\bibinfo {volume}
  {89}},\ \bibinfo {pages} {235127} (\bibinfo {year} {2014})}\BibitemShut
  {NoStop}%
\bibitem [{\citenamefont {Fang}\ \emph {et~al.}(2015)\citenamefont {Fang},
  \citenamefont {Chen}, \citenamefont {Kee},\ and\ \citenamefont
  {Fu}}]{fang2015topological}%
  \BibitemOpen
  \bibfield  {author} {\bibinfo {author} {\bibfnamefont {C.}~\bibnamefont
  {Fang}}, \bibinfo {author} {\bibfnamefont {Y.}~\bibnamefont {Chen}}, \bibinfo
  {author} {\bibfnamefont {H.-Y.}\ \bibnamefont {Kee}}, \ and\ \bibinfo
  {author} {\bibfnamefont {L.}~\bibnamefont {Fu}},\ }\bibfield  {title}
  {\enquote {\bibinfo {title} {Topological nodal line semimetals with and
  without spin-orbital coupling},}\ }\href@noop {} {\bibfield  {journal}
  {\bibinfo  {journal} {Phys. Rev. B}\ }\textbf {\bibinfo {volume} {92}},\
  \bibinfo {pages} {081201(R)} (\bibinfo {year} {2015})}\BibitemShut {NoStop}%
\bibitem [{\citenamefont {Zhao}\ and\ \citenamefont {Lu}(2017)}]{zhao2017pt}%
  \BibitemOpen
  \bibfield  {author} {\bibinfo {author} {\bibfnamefont {YX}~\bibnamefont
  {Zhao}}\ and\ \bibinfo {author} {\bibfnamefont {Y}~\bibnamefont {Lu}},\
  }\bibfield  {title} {\enquote {\bibinfo {title} {{$PT$}-symmetric real dirac
  fermions and semimetals},}\ }\href@noop {} {\bibfield  {journal} {\bibinfo
  {journal} {Phys. Rev. Lett.}\ }\textbf {\bibinfo {volume} {118}},\ \bibinfo
  {pages} {056401} (\bibinfo {year} {2017})}\BibitemShut {NoStop}%
\bibitem [{\citenamefont {Benalcazar}\ \emph
  {et~al.}(2017{\natexlab{a}})\citenamefont {Benalcazar}, \citenamefont
  {Bernevig},\ and\ \citenamefont {Hughes}}]{benalcazar2017quantized}%
  \BibitemOpen
  \bibfield  {author} {\bibinfo {author} {\bibfnamefont {W.~A.}\ \bibnamefont
  {Benalcazar}}, \bibinfo {author} {\bibfnamefont {B.~A.}\ \bibnamefont
  {Bernevig}}, \ and\ \bibinfo {author} {\bibfnamefont {T.~L.}\ \bibnamefont
  {Hughes}},\ }\bibfield  {title} {\enquote {\bibinfo {title} {Quantized
  electric multipole insulators},}\ }\href@noop {} {\bibfield  {journal}
  {\bibinfo  {journal} {Science}\ }\textbf {\bibinfo {volume} {357}},\ \bibinfo
  {pages} {61--66} (\bibinfo {year} {2017}{\natexlab{a}})}\BibitemShut
  {NoStop}%
\bibitem [{\citenamefont {Po}\ \emph {et~al.}(2018)\citenamefont {Po},
  \citenamefont {Watanabe},\ and\ \citenamefont {Vishwanath}}]{po2018fragile}%
  \BibitemOpen
  \bibfield  {author} {\bibinfo {author} {\bibfnamefont {H.~C.}\ \bibnamefont
  {Po}}, \bibinfo {author} {\bibfnamefont {H.}~\bibnamefont {Watanabe}}, \ and\
  \bibinfo {author} {\bibfnamefont {A.}~\bibnamefont {Vishwanath}},\ }\bibfield
   {title} {\enquote {\bibinfo {title} {Fragile topology and wannier
  obstructions},}\ }\href@noop {} {\bibfield  {journal} {\bibinfo  {journal}
  {Phys. Rev. Lett.}\ }\textbf {\bibinfo {volume} {121}},\ \bibinfo {pages}
  {126402} (\bibinfo {year} {2018})}\BibitemShut {NoStop}%
\bibitem [{\citenamefont {Cano}\ \emph {et~al.}(2018)\citenamefont {Cano},
  \citenamefont {Bradlyn}, \citenamefont {Wang}, \citenamefont {Elcoro},
  \citenamefont {Vergniory}, \citenamefont {Felser}, \citenamefont {Aroyo},\
  and\ \citenamefont {Bernevig}}]{cano2018topology}%
  \BibitemOpen
  \bibfield  {author} {\bibinfo {author} {\bibfnamefont {J.}~\bibnamefont
  {Cano}}, \bibinfo {author} {\bibfnamefont {B.}~\bibnamefont {Bradlyn}},
  \bibinfo {author} {\bibfnamefont {Z.}~\bibnamefont {Wang}}, \bibinfo {author}
  {\bibfnamefont {L.}~\bibnamefont {Elcoro}}, \bibinfo {author} {\bibfnamefont
  {M.G.}\ \bibnamefont {Vergniory}}, \bibinfo {author} {\bibfnamefont
  {C.}~\bibnamefont {Felser}}, \bibinfo {author} {\bibfnamefont {M.I.}\
  \bibnamefont {Aroyo}}, \ and\ \bibinfo {author} {\bibfnamefont {B.~A.}\
  \bibnamefont {Bernevig}},\ }\bibfield  {title} {\enquote {\bibinfo {title}
  {Topology of disconnected elementary band representations},}\ }\href@noop {}
  {\bibfield  {journal} {\bibinfo  {journal} {Phys. Rev. Lett.}\ }\textbf
  {\bibinfo {volume} {120}},\ \bibinfo {pages} {266401} (\bibinfo {year}
  {2018})}\BibitemShut {NoStop}%
\bibitem [{\citenamefont {Bouhon}\ \emph {et~al.}(2018)\citenamefont {Bouhon},
  \citenamefont {Black-Schaffer},\ and\ \citenamefont
  {Slager}}]{bouhon2018wilson}%
  \BibitemOpen
  \bibfield  {author} {\bibinfo {author} {\bibfnamefont {A.}~\bibnamefont
  {Bouhon}}, \bibinfo {author} {\bibfnamefont {A.~M.}\ \bibnamefont
  {Black-Schaffer}}, \ and\ \bibinfo {author} {\bibfnamefont {R.-J.}\
  \bibnamefont {Slager}},\ }\bibfield  {title} {\enquote {\bibinfo {title}
  {Wilson loop approach to topological crystalline insulators with time
  reversal symmetry},}\ }\href@noop {} {\bibfield  {journal} {\bibinfo
  {journal} {arXiv preprint arXiv:1804.09719}\ } (\bibinfo {year}
  {2018})}\BibitemShut {NoStop}%
\bibitem [{\citenamefont {Wang}\ \emph {et~al.}(2018)\citenamefont {Wang},
  \citenamefont {Wieder}, \citenamefont {Li}, \citenamefont {Yan},\ and\
  \citenamefont {Bernevig}}]{wang2018higher}%
  \BibitemOpen
  \bibfield  {author} {\bibinfo {author} {\bibfnamefont {Z.}~\bibnamefont
  {Wang}}, \bibinfo {author} {\bibfnamefont {B.~J.}\ \bibnamefont {Wieder}},
  \bibinfo {author} {\bibfnamefont {J.}~\bibnamefont {Li}}, \bibinfo {author}
  {\bibfnamefont {B.}~\bibnamefont {Yan}}, \ and\ \bibinfo {author}
  {\bibfnamefont {B.~A.}\ \bibnamefont {Bernevig}},\ }\bibfield  {title}
  {\enquote {\bibinfo {title} {Higher-order topology, monopole nodal lines, and
  the origin of large fermi arcs in transition metal dichalcogenides {XTe$_2$}
  (x= mo, w)},}\ }\href@noop {} {\bibfield  {journal} {\bibinfo  {journal}
  {arXiv preprint arXiv:1806.11116}\ } (\bibinfo {year} {2018})}\BibitemShut
  {NoStop}%
\bibitem [{\citenamefont {Bradlyn}\ \emph {et~al.}(2018)\citenamefont
  {Bradlyn}, \citenamefont {Wang}, \citenamefont {Cano},\ and\ \citenamefont
  {Bernevig}}]{bradlyn2018disconnected}%
  \BibitemOpen
  \bibfield  {author} {\bibinfo {author} {\bibfnamefont {B.}~\bibnamefont
  {Bradlyn}}, \bibinfo {author} {\bibfnamefont {Z.}~\bibnamefont {Wang}},
  \bibinfo {author} {\bibfnamefont {J.}~\bibnamefont {Cano}}, \ and\ \bibinfo
  {author} {\bibfnamefont {B.~A.}\ \bibnamefont {Bernevig}},\ }\bibfield
  {title} {\enquote {\bibinfo {title} {Disconnected elementary band
  representations, fragile topology, and wilson loops as topological
  indices},}\ }\href@noop {} {\bibfield  {journal} {\bibinfo  {journal}
  {arXiv:1807.09729}\ } (\bibinfo {year} {2018})}\BibitemShut {NoStop}%
\bibitem [{\citenamefont {Ahn}\ \emph {et~al.}(2018{\natexlab{b}})\citenamefont
  {Ahn}, \citenamefont {Park},\ and\ \citenamefont {Yang}}]{ahn2019failure}%
  \BibitemOpen
  \bibfield  {author} {\bibinfo {author} {\bibfnamefont {J.}~\bibnamefont
  {Ahn}}, \bibinfo {author} {\bibfnamefont {S.}~\bibnamefont {Park}}, \ and\
  \bibinfo {author} {\bibfnamefont {B.-J.}\ \bibnamefont {Yang}},\ }\bibfield
  {title} {\enquote {\bibinfo {title} {Failure of Nielsen-Ninomiya theorem and
  fragile topology in two-dimensional systems with space-time inversion
  symmetry: Application to twisted bilayer graphene at magic angle},}\
  }\href@noop {} {\bibfield  {journal} {\bibinfo
  {journal} {Phys. Rev. X}\ }\textbf {\bibinfo {volume} {9}},\ \bibinfo
  {pages} {021013} (\bibinfo {year} {2019})}\BibitemShut
  {NoStop}%
\bibitem [{\citenamefont {Wang}\ \emph {et~al.}(2010)\citenamefont {Wang},
  \citenamefont {Qi},\ and\ \citenamefont {Zhang}}]{wang2010equivalent}%
  \BibitemOpen
  \bibfield  {author} {\bibinfo {author} {\bibfnamefont {Z.}~\bibnamefont
  {Wang}}, \bibinfo {author} {\bibfnamefont {X.-L.}\ \bibnamefont {Qi}}, \ and\
  \bibinfo {author} {\bibfnamefont {S.-C.}\ \bibnamefont {Zhang}},\ }\bibfield
  {title} {\enquote {\bibinfo {title} {Equivalent topological invariants of
  topological insulators},}\ }\href@noop {} {\bibfield  {journal} {\bibinfo
  {journal} {New. J. Phys.}\ }\textbf {\bibinfo {volume} {12}},\
  \bibinfo {pages} {065007} (\bibinfo {year} {2010})}\BibitemShut {NoStop}%
\bibitem [{\citenamefont {Hughes}\ \emph {et~al.}(2011)\citenamefont {Hughes},
  \citenamefont {Prodan},\ and\ \citenamefont
  {Bernevig}}]{hughes2011inversion}%
  \BibitemOpen
  \bibfield  {author} {\bibinfo {author} {\bibfnamefont {T.~L.}\ \bibnamefont
  {Hughes}}, \bibinfo {author} {\bibfnamefont {E.}~\bibnamefont {Prodan}}, \
  and\ \bibinfo {author} {\bibfnamefont {B.~A.}\ \bibnamefont {Bernevig}},\
  }\bibfield  {title} {\enquote {\bibinfo {title} {Inversion-symmetric
  topological insulators},}\ }\href@noop {} {\bibfield  {journal} {\bibinfo
  {journal} {Phys. Rev. B}\ }\textbf {\bibinfo {volume} {83}},\ \bibinfo
  {pages} {245132} (\bibinfo {year} {2011})}\BibitemShut {NoStop}%
\bibitem [{\citenamefont {Bradlyn}\ \emph {et~al.}(2017)\citenamefont
  {Bradlyn}, \citenamefont {Elcoro}, \citenamefont {Cano}, \citenamefont
  {Vergniory}, \citenamefont {Wang}, \citenamefont {Felser}, \citenamefont
  {Aroyo},\ and\ \citenamefont {Bernevig}}]{bradlyn2017topological}%
  \BibitemOpen
  \bibfield  {author} {\bibinfo {author} {\bibfnamefont {B.}~\bibnamefont
  {Bradlyn}}, \bibinfo {author} {\bibfnamefont {L}~\bibnamefont {Elcoro}},
  \bibinfo {author} {\bibfnamefont {J.}~\bibnamefont {Cano}}, \bibinfo {author}
  {\bibfnamefont {MG}~\bibnamefont {Vergniory}}, \bibinfo {author}
  {\bibfnamefont {Z.}~\bibnamefont {Wang}}, \bibinfo {author} {\bibfnamefont
  {C}~\bibnamefont {Felser}}, \bibinfo {author} {\bibfnamefont
  {MI}~\bibnamefont {Aroyo}}, \ and\ \bibinfo {author} {\bibfnamefont {B.~A.}\
  \bibnamefont {Bernevig}},\ }\bibfield  {title} {\enquote {\bibinfo {title}
  {Topological quantum chemistry},}\ }\href@noop {} {\bibfield  {journal}
  {\bibinfo  {journal} {Nature}\ }\textbf {\bibinfo {volume} {547}},\ \bibinfo
  {pages} {298} (\bibinfo {year} {2017})}\BibitemShut {NoStop}%
\bibitem [{\citenamefont {Alexandradinata}\ and\ \citenamefont
  {H{\"o}ller}(2018)}]{alexandradinata2018no}%
  \BibitemOpen
  \bibfield  {author} {\bibinfo {author} {\bibfnamefont {A.}~\bibnamefont
  {Alexandradinata}}\ and\ \bibinfo {author} {\bibfnamefont {J.}~\bibnamefont
  {H{\"o}ller}},\ }\bibfield  {title} {\enquote {\bibinfo {title} {No-go
  theorem for topological insulators and high-throughput identification of
  chern insulators},}\ }\href@noop {} {\bibfield  {journal} {\bibinfo
  {journal} {Phys. Rev. B}\ }\textbf {\bibinfo {volume} {98}},\ \bibinfo
  {pages} {184305} (\bibinfo {year} {2018})}\BibitemShut {NoStop}%
\bibitem [{\citenamefont {Geier}\ \emph {et~al.}(2018)\citenamefont {Geier},
  \citenamefont {Trifunovic}, \citenamefont {Hoskam},\ and\ \citenamefont
  {Brouwer}}]{geier2018second}%
  \BibitemOpen
  \bibfield  {author} {\bibinfo {author} {\bibfnamefont {M.}~\bibnamefont
  {Geier}}, \bibinfo {author} {\bibfnamefont {L.}~\bibnamefont {Trifunovic}},
  \bibinfo {author} {\bibfnamefont {M.}~\bibnamefont {Hoskam}}, \ and\ \bibinfo
  {author} {\bibfnamefont {P.~W.}\ \bibnamefont {Brouwer}},\ }\bibfield
  {title} {\enquote {\bibinfo {title} {Second-order topological insulators and
  superconductors with an order-two crystalline symmetry},}\ }\href@noop {}
  {\bibfield  {journal} {\bibinfo  {journal} {Phys. Rev. B}\ }\textbf {\bibinfo
  {volume} {97}},\ \bibinfo {pages} {205135} (\bibinfo {year}
  {2018})}\BibitemShut {NoStop}%
\bibitem [{\citenamefont {Trifunovic}\ and\ \citenamefont
  {Brouwer}(2019)}]{trifunovic2019higher}%
  \BibitemOpen
  \bibfield  {author} {\bibinfo {author} {\bibfnamefont {L.}~\bibnamefont
  {Trifunovic}}\ and\ \bibinfo {author} {\bibfnamefont {P.~W.}\ \bibnamefont
  {Brouwer}},\ }\bibfield  {title} {\enquote {\bibinfo {title} {Higher-order
  bulk-boundary correspondence for topological crystalline phases},}\
  }\href@noop {} {\bibfield  {journal} {\bibinfo  {journal} {Phys. Rev. X}\
  }\textbf {\bibinfo {volume} {9}},\ \bibinfo {pages} {011012} (\bibinfo {year}
  {2019})}\BibitemShut {NoStop}%
\bibitem [{\citenamefont {Benalcazar}\ \emph
  {et~al.}(2017{\natexlab{b}})\citenamefont {Benalcazar}, \citenamefont
  {Bernevig},\ and\ \citenamefont {Hughes}}]{benalcazar2017electric}%
  \BibitemOpen
  \bibfield  {author} {\bibinfo {author} {\bibfnamefont {W.~A.}\ \bibnamefont
  {Benalcazar}}, \bibinfo {author} {\bibfnamefont {B.~A.}\ \bibnamefont
  {Bernevig}}, \ and\ \bibinfo {author} {\bibfnamefont {T.~L.}\ \bibnamefont
  {Hughes}},\ }\bibfield  {title} {\enquote {\bibinfo {title} {Electric
  multipole moments, topological multipole moment pumping, and chiral hinge
  states in crystalline insulators},}\ }\href@noop {} {\bibfield  {journal}
  {\bibinfo  {journal} {Phys. Rev. B}\ }\textbf {\bibinfo {volume} {96}},\
  \bibinfo {pages} {245115} (\bibinfo {year} {2017}{\natexlab{b}})}\BibitemShut
  {NoStop}%
\bibitem [{\citenamefont {Serra-Garcia}\ \emph {et~al.}(2018)\citenamefont
  {Serra-Garcia}, \citenamefont {Peri}, \citenamefont {S{\"u}sstrunk},
  \citenamefont {Bilal}, \citenamefont {Larsen}, \citenamefont {Villanueva},\
  and\ \citenamefont {Huber}}]{serra2018observation}%
  \BibitemOpen
  \bibfield  {author} {\bibinfo {author} {\bibfnamefont {M.}~\bibnamefont
  {Serra-Garcia}}, \bibinfo {author} {\bibfnamefont {V.}~\bibnamefont {Peri}},
  \bibinfo {author} {\bibfnamefont {R.}~\bibnamefont {S{\"u}sstrunk}}, \bibinfo
  {author} {\bibfnamefont {O.~R.}\ \bibnamefont {Bilal}}, \bibinfo {author}
  {\bibfnamefont {T.}~\bibnamefont {Larsen}}, \bibinfo {author} {\bibfnamefont
  {L.~G.}\ \bibnamefont {Villanueva}}, \ and\ \bibinfo {author} {\bibfnamefont
  {S.~D.}\ \bibnamefont {Huber}},\ }\bibfield  {title} {\enquote {\bibinfo
  {title} {Observation of a phononic quadrupole topological insulator},}\
  }\href@noop {} {\bibfield  {journal} {\bibinfo  {journal} {Nature}\ }\textbf
  {\bibinfo {volume} {555}},\ \bibinfo {pages} {342} (\bibinfo {year}
  {2018})}\BibitemShut {NoStop}%
\bibitem [{\citenamefont {Imhof}\ \emph {et~al.}(2018)\citenamefont {Imhof},
  \citenamefont {Berger}, \citenamefont {Bayer}, \citenamefont {Brehm},
  \citenamefont {Molenkamp}, \citenamefont {Kiessling}, \citenamefont
  {Schindler}, \citenamefont {Lee}, \citenamefont {Greiter}, \citenamefont
  {Neupert} \emph {et~al.}}]{imhof2018topolectrical}%
  \BibitemOpen
  \bibfield  {author} {\bibinfo {author} {\bibfnamefont {S.}~\bibnamefont
  {Imhof}}, \bibinfo {author} {\bibfnamefont {C.}~\bibnamefont {Berger}},
  \bibinfo {author} {\bibfnamefont {F.}~\bibnamefont {Bayer}}, \bibinfo
  {author} {\bibfnamefont {J.}~\bibnamefont {Brehm}}, \bibinfo {author}
  {\bibfnamefont {L.~W}\ \bibnamefont {Molenkamp}}, \bibinfo {author}
  {\bibfnamefont {T.}~\bibnamefont {Kiessling}}, \bibinfo {author}
  {\bibfnamefont {F.}~\bibnamefont {Schindler}}, \bibinfo {author}
  {\bibfnamefont {C.~H.}\ \bibnamefont {Lee}}, \bibinfo {author} {\bibfnamefont
  {M.}~\bibnamefont {Greiter}}, \bibinfo {author} {\bibfnamefont
  {T.}~\bibnamefont {Neupert}},  \emph {et~al.},\ }\bibfield  {title} {\enquote
  {\bibinfo {title} {Topolectrical-circuit realization of topological corner
  modes},}\ }\href@noop {} {\bibfield  {journal} {\bibinfo  {journal} {Nat.
  Phys.}\ }\textbf {\bibinfo {volume} {14}},\ \bibinfo {pages} {925} (\bibinfo
  {year} {2018})}\BibitemShut {NoStop}%
\bibitem [{\citenamefont {Langbehn}\ \emph {et~al.}(2017)\citenamefont
  {Langbehn}, \citenamefont {Peng}, \citenamefont {Trifunovic}, \citenamefont
  {von Oppen},\ and\ \citenamefont {Brouwer}}]{langbehn2017reflection}%
  \BibitemOpen
  \bibfield  {author} {\bibinfo {author} {\bibfnamefont {J.}~\bibnamefont
  {Langbehn}}, \bibinfo {author} {\bibfnamefont {Y.}~\bibnamefont {Peng}},
  \bibinfo {author} {\bibfnamefont {L.}~\bibnamefont {Trifunovic}}, \bibinfo
  {author} {\bibfnamefont {F.}~\bibnamefont {von Oppen}}, \ and\ \bibinfo
  {author} {\bibfnamefont {P.~W.}\ \bibnamefont {Brouwer}},\ }\bibfield
  {title} {\enquote {\bibinfo {title} {Reflection-symmetric second-order
  topological insulators and superconductors},}\ }\href@noop {} {\bibfield
  {journal} {\bibinfo  {journal} {Phys. Rev. Lett.}\ }\textbf {\bibinfo
  {volume} {119}},\ \bibinfo {pages} {246401} (\bibinfo {year}
  {2017})}\BibitemShut {NoStop}%
\bibitem [{\citenamefont {Song}\ \emph {et~al.}(2017)\citenamefont {Song},
  \citenamefont {Fang},\ and\ \citenamefont {Fang}}]{song2017d}%
  \BibitemOpen
  \bibfield  {author} {\bibinfo {author} {\bibfnamefont {Z.}~\bibnamefont
  {Song}}, \bibinfo {author} {\bibfnamefont {Z.}~\bibnamefont {Fang}}, \ and\
  \bibinfo {author} {\bibfnamefont {C.}~\bibnamefont {Fang}},\ }\bibfield
  {title} {\enquote {\bibinfo {title} {($d- 2$)-dimensional edge states of
  rotation symmetry protected topological states},}\ }\href@noop {} {\bibfield
  {journal} {\bibinfo  {journal} {Phys. Rev. Lett.}\ }\textbf {\bibinfo
  {volume} {119}},\ \bibinfo {pages} {246402} (\bibinfo {year}
  {2017})}\BibitemShut {NoStop}%
\bibitem [{\citenamefont {Schindler}\ \emph
  {et~al.}(2018{\natexlab{a}})\citenamefont {Schindler}, \citenamefont {Wang},
  \citenamefont {Vergniory}, \citenamefont {Cook}, \citenamefont {Murani},
  \citenamefont {Sengupta}, \citenamefont {Kasumov}, \citenamefont {Deblock},
  \citenamefont {Jeon}, \citenamefont {Drozdov} \emph
  {et~al.}}]{schindler2018bismuth}%
  \BibitemOpen
  \bibfield  {author} {\bibinfo {author} {\bibfnamefont {F.}~\bibnamefont
  {Schindler}}, \bibinfo {author} {\bibfnamefont {Z.}~\bibnamefont {Wang}},
  \bibinfo {author} {\bibfnamefont {M.~G.}\ \bibnamefont {Vergniory}}, \bibinfo
  {author} {\bibfnamefont {A.~M.}\ \bibnamefont {Cook}}, \bibinfo {author}
  {\bibfnamefont {A.}~\bibnamefont {Murani}}, \bibinfo {author} {\bibfnamefont
  {S.}~\bibnamefont {Sengupta}}, \bibinfo {author} {\bibfnamefont {A.~Y.}\
  \bibnamefont {Kasumov}}, \bibinfo {author} {\bibfnamefont {R.}~\bibnamefont
  {Deblock}}, \bibinfo {author} {\bibfnamefont {S.}~\bibnamefont {Jeon}},
  \bibinfo {author} {\bibfnamefont {I.}~\bibnamefont {Drozdov}},  \emph
  {et~al.},\ }\bibfield  {title} {\enquote {\bibinfo {title} {Higher-order
  topology in bismuth},}\ }\href@noop {} {\bibfield  {journal} {\bibinfo
  {journal} {Nat. Phys.}\ }\textbf {\bibinfo {volume} {14}},\ \bibinfo {pages}
  {918} (\bibinfo {year} {2018}{\natexlab{a}})}\BibitemShut {NoStop}%
\bibitem [{\citenamefont {Matsugatani}\ and\ \citenamefont
  {Watanabe}(2018)}]{matsugatani2018connecting}%
  \BibitemOpen
  \bibfield  {author} {\bibinfo {author} {\bibfnamefont {A.}~\bibnamefont
  {Matsugatani}}\ and\ \bibinfo {author} {\bibfnamefont {H.}~\bibnamefont
  {Watanabe}},\ }\bibfield  {title} {\enquote {\bibinfo {title} {Connecting
  higher-order topological insulators to lower-dimensional topological
  insulators},}\ }\href@noop {} {\bibfield  {journal} {\bibinfo  {journal}
  {arXiv:1804.02794}\ } (\bibinfo {year} {2018})}\BibitemShut {NoStop}%
\bibitem [{\citenamefont {Franca}\ \emph {et~al.}(2018)\citenamefont {Franca},
  \citenamefont {Brink},\ and\ \citenamefont {Fulga}}]{franca2018anomalous}%
  \BibitemOpen
  \bibfield  {author} {\bibinfo {author} {\bibfnamefont {S.}~\bibnamefont
  {Franca}}, \bibinfo {author} {\bibfnamefont {J.}~\bibnamefont {Brink}}, \
  and\ \bibinfo {author} {\bibfnamefont {I.~C.}\ \bibnamefont {Fulga}},\
  }\bibfield  {title} {\enquote {\bibinfo {title} {Anomalous higher-order
  topological insulators},}\ }\href@noop {} {\bibfield  {journal} {\bibinfo
  {journal} {arXiv:1807.09050}\ } (\bibinfo {year} {2018})}\BibitemShut
  {NoStop}%
\bibitem [{\citenamefont {Calugaru}\ \emph {et~al.}(2018)\citenamefont
  {Calugaru}, \citenamefont {Juricic},\ and\ \citenamefont
  {Roy}}]{calugaru2018higher}%
  \BibitemOpen
  \bibfield  {author} {\bibinfo {author} {\bibfnamefont {D.}~\bibnamefont
  {Calugaru}}, \bibinfo {author} {\bibfnamefont {V.}~\bibnamefont {Juricic}}, \
  and\ \bibinfo {author} {\bibfnamefont {B.}~\bibnamefont {Roy}},\ }\bibfield
  {title} {\enquote {\bibinfo {title} {Higher order topological phases: A
  general principle of construction},}\ }\href@noop {} {\bibfield  {journal}
  {\bibinfo  {journal} {arXiv:1808.08965}\ } (\bibinfo {year}
  {2018})}\BibitemShut {NoStop}%
\bibitem [{\citenamefont {Benalcazar}\ \emph {et~al.}(2018)\citenamefont
  {Benalcazar}, \citenamefont {Li},\ and\ \citenamefont
  {Hughes}}]{benalcazar2018quantization}%
  \BibitemOpen
  \bibfield  {author} {\bibinfo {author} {\bibfnamefont {W.~A.}\ \bibnamefont
  {Benalcazar}}, \bibinfo {author} {\bibfnamefont {T.}~\bibnamefont {Li}}, \
  and\ \bibinfo {author} {\bibfnamefont {T.~L.}\ \bibnamefont {Hughes}},\
  }\bibfield  {title} {\enquote {\bibinfo {title} {Quantization of fractional
  corner charge in $C_n$-symmetric topological crystalline insulators},}\
  }\href@noop {} {\bibfield  {journal} {\bibinfo  {journal} {arXiv:1809.02142}\
  } (\bibinfo {year} {2018})}\BibitemShut {NoStop}%
\bibitem [{\citenamefont {Ezawa}(2018{\natexlab{a}})}]{ezawa2018higher}%
  \BibitemOpen
  \bibfield  {author} {\bibinfo {author} {\bibfnamefont {M.}~\bibnamefont
  {Ezawa}},\ }\bibfield  {title} {\enquote {\bibinfo {title} {Higher-order
  topological insulators and semimetals on the breathing kagome and pyrochlore
  lattices},}\ }\href@noop {} {\bibfield  {journal} {\bibinfo  {journal} {Phys.
  Rev. Lett.}\ }\textbf {\bibinfo {volume} {120}},\ \bibinfo {pages} {026801}
  (\bibinfo {year} {2018}{\natexlab{a}})}\BibitemShut {NoStop}%
\bibitem [{\citenamefont {Ezawa}(2018{\natexlab{b}})}]{ezawa2018simple}%
  \BibitemOpen
  \bibfield  {author} {\bibinfo {author} {\bibfnamefont {M.}~\bibnamefont
  {Ezawa}},\ }\bibfield  {title} {\enquote {\bibinfo {title} {Simple model for
  second-order topological insulators and loop-nodal semimetals in transition
  metal dichalcogenides {XTe$_2 $(X= Mo, W)}},}\ }\href@noop {} {\bibfield
  {journal} {\bibinfo  {journal} {arXiv:1807.10932}\ } (\bibinfo {year}
  {2018}{\natexlab{b}})}\BibitemShut {NoStop}%
\bibitem [{\citenamefont {Zhang}\ \emph {et~al.}(2013)\citenamefont {Zhang},
  \citenamefont {Kane},\ and\ \citenamefont {Mele}}]{zhang2013surface}%
  \BibitemOpen
  \bibfield  {author} {\bibinfo {author} {\bibfnamefont {F.}~\bibnamefont
  {Zhang}}, \bibinfo {author} {\bibfnamefont {C.~L.}\ \bibnamefont {Kane}}, \
  and\ \bibinfo {author} {\bibfnamefont {E.~J.}\ \bibnamefont {Mele}},\
  }\bibfield  {title} {\enquote {\bibinfo {title} {Surface state magnetization
  and chiral edge states on topological insulators},}\ }\href@noop {}
  {\bibfield  {journal} {\bibinfo  {journal} {Phys. Rev. Lett.}\ }\textbf
  {\bibinfo {volume} {110}},\ \bibinfo {pages} {046404} (\bibinfo {year}
  {2013})}\BibitemShut {NoStop}%
\bibitem [{\citenamefont {Fang}\ and\ \citenamefont
  {Fu}(2017)}]{fang2017rotation}%
  \BibitemOpen
  \bibfield  {author} {\bibinfo {author} {\bibfnamefont {C.}~\bibnamefont
  {Fang}}\ and\ \bibinfo {author} {\bibfnamefont {L.}~\bibnamefont {Fu}},\
  }\bibfield  {title} {\enquote {\bibinfo {title} {Rotation anomaly and
  topological crystalline insulators},}\ }\href@noop {} {\bibfield  {journal}
  {\bibinfo  {journal} {arXiv preprint arXiv:1709.01929}\ } (\bibinfo {year}
  {2017})}\BibitemShut {NoStop}%
\bibitem [{\citenamefont {Khalaf}(2018)}]{khalaf2018higher}%
  \BibitemOpen
  \bibfield  {author} {\bibinfo {author} {\bibfnamefont {E.}~\bibnamefont
  {Khalaf}},\ }\bibfield  {title} {\enquote {\bibinfo {title} {Higher-order
  topological insulators and superconductors protected by inversion
  symmetry},}\ }\href@noop {} {\bibfield  {journal} {\bibinfo  {journal} {Phys.
  Rev. B}\ }\textbf {\bibinfo {volume} {97}},\ \bibinfo {pages} {205136}
  (\bibinfo {year} {2018})}\BibitemShut {NoStop}%
\bibitem [{\citenamefont {Kooi}\ \emph {et~al.}(2018)\citenamefont {Kooi},
  \citenamefont {vanMiert},\ and\ \citenamefont {Ortix}}]{kooi2018inversion}%
  \BibitemOpen
  \bibfield  {author} {\bibinfo {author} {\bibfnamefont {S.~H.}\ \bibnamefont
  {Kooi}}, \bibinfo {author} {\bibfnamefont {G.}~\bibnamefont {vanMiert}}, \
  and\ \bibinfo {author} {\bibfnamefont {C.}~\bibnamefont {Ortix}},\ }\bibfield
   {title} {\enquote {\bibinfo {title} {Inversion-symmetry protected chiral
  hinge states in stacks of doped quantum hall layers},}\ }\href@noop {}
  {\bibfield  {journal} {\bibinfo  {journal} {arXiv:1807.01277}\ } (\bibinfo
  {year} {2018})}\BibitemShut {NoStop}%
\bibitem [{\citenamefont {Varnava}\ and\ \citenamefont
  {Vanderbilt}(2018)}]{varnava2018surfaces}%
  \BibitemOpen
  \bibfield  {author} {\bibinfo {author} {\bibfnamefont {N.}~\bibnamefont
  {Varnava}}\ and\ \bibinfo {author} {\bibfnamefont {D.}~\bibnamefont
  {Vanderbilt}},\ }\bibfield  {title} {\enquote {\bibinfo {title} {Surfaces of
  axion insulators},}\ }\href@noop {} {\bibfield  {journal} {\bibinfo
  {journal} {arXiv:1809.02853}\ } (\bibinfo {year} {2018})}\BibitemShut
  {NoStop}%
\bibitem [{\citenamefont {vanMiert}\ and\ \citenamefont
  {Ortix}(2018)}]{vanmiert2018higher}%
  \BibitemOpen
  \bibfield  {author} {\bibinfo {author} {\bibfnamefont {Guido}\ \bibnamefont
  {vanMiert}}\ and\ \bibinfo {author} {\bibfnamefont {Carmine}\ \bibnamefont
  {Ortix}},\ }\bibfield  {title} {\enquote {\bibinfo {title} {Higher-order
  topological insulators protected by inversion and rotoinversion
  symmetries},}\ }\href@noop {} {\bibfield  {journal} {\bibinfo  {journal}
  {Phys. Rev. B}\ }\textbf {\bibinfo {volume} {98}},\ \bibinfo {pages}
  {081110(R)} (\bibinfo {year} {2018})}\BibitemShut {NoStop}%
\bibitem [{\citenamefont {Yue}\ \emph {et~al.}(2019)\citenamefont {Yue},
  \citenamefont {Xu}, \citenamefont {Song}, \citenamefont {Weng}, \citenamefont
  {Lu}, \citenamefont {Fang},\ and\ \citenamefont {Dai}}]{yue2019symmetry}%
  \BibitemOpen
  \bibfield  {author} {\bibinfo {author} {\bibfnamefont {C.}~\bibnamefont
  {Yue}}, \bibinfo {author} {\bibfnamefont {Y.}~\bibnamefont {Xu}}, \bibinfo
  {author} {\bibfnamefont {Z.}~\bibnamefont {Song}}, \bibinfo {author}
  {\bibfnamefont {H.}~\bibnamefont {Weng}}, \bibinfo {author} {\bibfnamefont
  {Y.-M.}\ \bibnamefont {Lu}}, \bibinfo {author} {\bibfnamefont
  {C.}~\bibnamefont {Fang}}, \ and\ \bibinfo {author} {\bibfnamefont
  {X.}~\bibnamefont {Dai}},\ }\bibfield  {title} {\enquote {\bibinfo {title}
  {Symmetry-enforced chiral hinge states and surface quantum anomalous hall
  effect in the magnetic axion insulator {Bi$_{2-x}$Sm$_x$Se$_3$}},}\
  }\href@noop {} {\bibfield  {journal} {\bibinfo  {journal} {Nat. Phys.}\textbf{\bibinfo {volume} {15}}
  ,\ \bibinfo {pages} {571}} (\bibinfo {year} {2019})}\BibitemShut {NoStop}%
\bibitem [{\citenamefont {Schindler}\ \emph
  {et~al.}(2018{\natexlab{b}})\citenamefont {Schindler}, \citenamefont {Cook},
  \citenamefont {Vergniory}, \citenamefont {Wang}, \citenamefont {Parkin},
  \citenamefont {Bernevig},\ and\ \citenamefont
  {Neupert}}]{schindler2018higher}%
  \BibitemOpen
  \bibfield  {author} {\bibinfo {author} {\bibfnamefont {F.}~\bibnamefont
  {Schindler}}, \bibinfo {author} {\bibfnamefont {A.~M.}\ \bibnamefont {Cook}},
  \bibinfo {author} {\bibfnamefont {M.~G.}\ \bibnamefont {Vergniory}}, \bibinfo
  {author} {\bibfnamefont {Z.}~\bibnamefont {Wang}}, \bibinfo {author}
  {\bibfnamefont {S.~S.~P.}\ \bibnamefont {Parkin}}, \bibinfo {author}
  {\bibfnamefont {B.~A.}\ \bibnamefont {Bernevig}}, \ and\ \bibinfo {author}
  {\bibfnamefont {T.}~\bibnamefont {Neupert}},\ }\bibfield  {title} {\enquote
  {\bibinfo {title} {Higher-order topological insulators},}\ }\href@noop {}
  {\bibfield  {journal} {\bibinfo  {journal} {Sci. Adv.}\ }\textbf
  {\bibinfo {volume} {4}},\ \bibinfo {pages} {eaat0346} (\bibinfo {year}
  {2018}{\natexlab{b}})}\BibitemShut {NoStop}%
\bibitem [{\citenamefont {Ezawa}(2018{\natexlab{c}})}]{ezawa2018strong}%
  \BibitemOpen
  \bibfield  {author} {\bibinfo {author} {\bibfnamefont {M.}~\bibnamefont
  {Ezawa}},\ }\bibfield  {title} {\enquote {\bibinfo {title} {Strong and weak
  second-order topological insulators with hexagonal symmetry and {Z3}
  index},}\ }\href@noop {} {\bibfield  {journal} {\bibinfo  {journal} {Phys.
  Rev. B}\ }\textbf {\bibinfo {volume} {97}},\ \bibinfo {pages} {241402(R)}
  (\bibinfo {year} {2018}{\natexlab{c}})}\BibitemShut {NoStop}%
\bibitem [{\citenamefont {Ezawa}(2018{\natexlab{d}})}]{ezawa2018magnetic}%
  \BibitemOpen
  \bibfield  {author} {\bibinfo {author} {\bibfnamefont {M.}~\bibnamefont
  {Ezawa}},\ }\bibfield  {title} {\enquote {\bibinfo {title} {Magnetic
  second-order topological insulators and semimetals},}\ }\href@noop {}
  {\bibfield  {journal} {\bibinfo  {journal} {Phys. Rev. B}\ }\textbf {\bibinfo
  {volume} {97}},\ \bibinfo {pages} {155305} (\bibinfo {year}
  {2018}{\natexlab{d}})}\BibitemShut {NoStop}%
\bibitem [{\citenamefont {Turner}\ \emph {et~al.}(2012)\citenamefont {Turner},
  \citenamefont {Zhang}, \citenamefont {Mong},\ and\ \citenamefont
  {Vishwanath}}]{turner2012quantized}%
  \BibitemOpen
  \bibfield  {author} {\bibinfo {author} {\bibfnamefont {A.~M.}\ \bibnamefont
  {Turner}}, \bibinfo {author} {\bibfnamefont {Y.}~\bibnamefont {Zhang}},
  \bibinfo {author} {\bibfnamefont {R.~S.~K.}\ \bibnamefont {Mong}}, \ and\
  \bibinfo {author} {\bibfnamefont {A.}~\bibnamefont {Vishwanath}},\ }\bibfield
   {title} {\enquote {\bibinfo {title} {Quantized response and topology of
  magnetic insulators with inversion symmetry},}\ }\href@noop {} {\bibfield
  {journal} {\bibinfo  {journal} {Phys. Rev. B}\ }\textbf {\bibinfo {volume}
  {85}},\ \bibinfo {pages} {165120} (\bibinfo {year} {2012})}\BibitemShut
  {NoStop}%
\bibitem [{\citenamefont {Trifunovic}\ and\ \citenamefont
  {Brouwer}(2017)}]{trifunovic2017bott}%
  \BibitemOpen
  \bibfield  {author} {\bibinfo {author} {\bibfnamefont {L.}~\bibnamefont
  {Trifunovic}}\ and\ \bibinfo {author} {\bibfnamefont {P.}~\bibnamefont
  {Brouwer}},\ }\bibfield  {title} {\enquote {\bibinfo {title} {Bott
  periodicity for the topological classification of gapped states of matter
  with reflection symmetry},}\ }\href@noop {} {\bibfield  {journal} {\bibinfo
  {journal} {Phys. Rev. B}\ }\textbf {\bibinfo {volume} {96}},\ \bibinfo
  {pages} {195109} (\bibinfo {year} {2017})}\BibitemShut {NoStop}%
\bibitem [{\citenamefont {Sun}\ \emph {et~al.}(2018)\citenamefont {Sun},
  \citenamefont {Zhang},\ and\ \citenamefont
  {Bzdu{\v{s}}ek}}]{sun2018conversion}%
  \BibitemOpen
  \bibfield  {author} {\bibinfo {author} {\bibfnamefont {X.-Q.}\ \bibnamefont
  {Sun}}, \bibinfo {author} {\bibfnamefont {S.-C.}\ \bibnamefont {Zhang}}, \
  and\ \bibinfo {author} {\bibfnamefont {T.}~\bibnamefont {Bzdu{\v{s}}ek}},\
  }\bibfield  {title} {\enquote {\bibinfo {title} {Conversion rules for weyl
  points and nodal lines in topological media},}\ }\href@noop {} {\bibfield
  {journal} {\bibinfo  {journal} {Phys. Rev. Lett.}\ }\textbf {\bibinfo
  {volume} {121}},\ \bibinfo {pages} {106402} (\bibinfo {year}
  {2018})}\BibitemShut {NoStop}%
\bibitem [{\citenamefont {Bzdu{\v{s}}ek}\ and\ \citenamefont
  {Sigrist}(2017)}]{bzdusek2017robust}%
  \BibitemOpen
  \bibfield  {author} {\bibinfo {author} {\bibfnamefont {Tom{\'a}{\v{s}}}\
  \bibnamefont {Bzdu{\v{s}}ek}}\ and\ \bibinfo {author} {\bibfnamefont
  {Manfred}\ \bibnamefont {Sigrist}},\ }\bibfield  {title} {\enquote {\bibinfo
  {title} {Robust doubly charged nodal lines and nodal surfaces in
  centrosymmetric systems},}\ }\href@noop {} {\bibfield  {journal} {\bibinfo
  {journal} {Phys. Rev. B}\ }\textbf {\bibinfo {volume} {96}},\ \bibinfo
  {pages} {155105} (\bibinfo {year} {2017})}\BibitemShut {NoStop}%
\bibitem [{\citenamefont {Hatcher}(2002)}]{hatcher2002algebraic}%
  \BibitemOpen
  \bibfield  {author} {\bibinfo {author} {\bibfnamefont {A.}~\bibnamefont
  {Hatcher}},\ }\href@noop {} {\emph {\bibinfo {title} {Algebraic Topology}}}\
  (\bibinfo  {publisher} {Cambridge University Press},\ \bibinfo {year}
  {2002})\BibitemShut {NoStop}%
\bibitem [{\citenamefont {Hatcher}(unpublished)}]{hatcher2003vector}%
  \BibitemOpen
  \bibfield  {author} {\bibinfo {author} {\bibfnamefont {A.}~\bibnamefont
  {Hatcher}},\ }\bibfield  {title} {\enquote {\bibinfo {title} {Vector bundles
  and K-theory},}\ }\href@noop {} {\bibfield  {journal} {\bibinfo  {journal}
  {http://pi.math.cornell.edu/~hatcher/VBKT/VB.pdf}\ } (\bibinfo {year}
  {unpublished})}\BibitemShut {NoStop}%
\bibitem [{\citenamefont {Read}(2017)}]{read2017compactly}%
  \BibitemOpen
  \bibfield  {author} {\bibinfo {author} {\bibfnamefont {N.}~\bibnamefont
  {Read}},\ }\bibfield  {title} {\enquote {\bibinfo {title} {Compactly
  supported wannier functions and algebraic k-theory},}\ }\href@noop {}
  {\bibfield  {journal} {\bibinfo  {journal} {Phys. Rev. B}\ }\textbf {\bibinfo
  {volume} {95}},\ \bibinfo {pages} {115309} (\bibinfo {year}
  {2017})}\BibitemShut {NoStop}%
\bibitem [{\citenamefont {Brouder}\ \emph {et~al.}(2007)\citenamefont
  {Brouder}, \citenamefont {Panati}, \citenamefont {Calandra}, \citenamefont
  {Mourougane},\ and\ \citenamefont {Marzari}}]{brouder2007exponential}%
  \BibitemOpen
  \bibfield  {author} {\bibinfo {author} {\bibfnamefont {C.}~\bibnamefont
  {Brouder}}, \bibinfo {author} {\bibfnamefont {G.}~\bibnamefont {Panati}},
  \bibinfo {author} {\bibfnamefont {M.}~\bibnamefont {Calandra}}, \bibinfo
  {author} {\bibfnamefont {C.}~\bibnamefont {Mourougane}}, \ and\ \bibinfo
  {author} {\bibfnamefont {N.}~\bibnamefont {Marzari}},\ }\bibfield  {title}
  {\enquote {\bibinfo {title} {Exponential localization of wannier functions in
  insulators},}\ }\href@noop {} {\bibfield  {journal} {\bibinfo  {journal}
  {Phys. Rev. Lett.}\ }\textbf {\bibinfo {volume} {98}},\ \bibinfo {pages}
  {046402} (\bibinfo {year} {2007})}\BibitemShut {NoStop}%
\bibitem [{\citenamefont {Qi}\ \emph {et~al.}(2008)\citenamefont {Qi},
  \citenamefont {Hughes},\ and\ \citenamefont {Zhang}}]{qi2008topological}%
  \BibitemOpen
  \bibfield  {author} {\bibinfo {author} {\bibfnamefont {X.-L.}\ \bibnamefont
  {Qi}}, \bibinfo {author} {\bibfnamefont {T.~L.}\ \bibnamefont {Hughes}}, \
  and\ \bibinfo {author} {\bibfnamefont {S.-C.}\ \bibnamefont {Zhang}},\
  }\bibfield  {title} {\enquote {\bibinfo {title} {Topological field theory of
  time-reversal invariant insulators},}\ }\href@noop {} {\bibfield  {journal}
  {\bibinfo  {journal} {Phys. Rev. B}\ }\textbf {\bibinfo {volume} {78}},\
  \bibinfo {pages} {195424} (\bibinfo {year} {2008})}\BibitemShut {NoStop}%
\bibitem [{\citenamefont {Essin}\ \emph {et~al.}(2009)\citenamefont {Essin},
  \citenamefont {Moore},\ and\ \citenamefont
  {Vanderbilt}}]{essin2009magnetoelectric}%
  \BibitemOpen
  \bibfield  {author} {\bibinfo {author} {\bibfnamefont {A.~M.}\ \bibnamefont
  {Essin}}, \bibinfo {author} {\bibfnamefont {J.~E.}\ \bibnamefont {Moore}}, \
  and\ \bibinfo {author} {\bibfnamefont {D.}~\bibnamefont {Vanderbilt}},\
  }\bibfield  {title} {\enquote {\bibinfo {title} {Magnetoelectric
  polarizability and axion electrodynamics in crystalline insulators},}\
  }\href@noop {} {\bibfield  {journal} {\bibinfo  {journal} {Phys. Rev. Lett.}\
  }\textbf {\bibinfo {volume} {102}},\ \bibinfo {pages} {146805} (\bibinfo
  {year} {2009})}\BibitemShut {NoStop}%
\bibitem [{\citenamefont {Fu}\ \emph {et~al.}(2007)\citenamefont {Fu},
  \citenamefont {Kane},\ and\ \citenamefont {Mele}}]{fu2007topological}%
  \BibitemOpen
  \bibfield  {author} {\bibinfo {author} {\bibfnamefont {L.}~\bibnamefont
  {Fu}}, \bibinfo {author} {\bibfnamefont {C.~L.}\ \bibnamefont {Kane}}, \ and\
  \bibinfo {author} {\bibfnamefont {E.~J.}\ \bibnamefont {Mele}},\ }\bibfield
  {title} {\enquote {\bibinfo {title} {Topological insulators in three
  dimensions},}\ }\href@noop {} {\bibfield  {journal} {\bibinfo  {journal}
  {Phys. Rev. Lett.}\ }\textbf {\bibinfo {volume} {98}},\ \bibinfo {pages}
  {106803} (\bibinfo {year} {2007})}\BibitemShut {NoStop}%
\bibitem [{\citenamefont {Yu}\ \emph {et~al.}(2011)\citenamefont {Yu},
  \citenamefont {Qi}, \citenamefont {Bernevig}, \citenamefont {Fang},\ and\
  \citenamefont {Dai}}]{yu2011equivalent}%
  \BibitemOpen
  \bibfield  {author} {\bibinfo {author} {\bibfnamefont {R.}~\bibnamefont
  {Yu}}, \bibinfo {author} {\bibfnamefont {X.-L.}\ \bibnamefont {Qi}}, \bibinfo
  {author} {\bibfnamefont {A.}~\bibnamefont {Bernevig}}, \bibinfo {author}
  {\bibfnamefont {Z.}~\bibnamefont {Fang}}, \ and\ \bibinfo {author}
  {\bibfnamefont {X.}~\bibnamefont {Dai}},\ }\bibfield  {title} {\enquote
  {\bibinfo {title} {Equivalent expression of {$Z_2$} topological invariant for
  band insulators using the non-abelian berry connection},}\ }\href@noop {}
  {\bibfield  {journal} {\bibinfo  {journal} {Phys. Rev. B}\ }\textbf {\bibinfo
  {volume} {84}},\ \bibinfo {pages} {075119} (\bibinfo {year}
  {2011})}\BibitemShut {NoStop}%
\bibitem [{\citenamefont {Song}\ \emph {et~al.}(2018)\citenamefont {Song},
  \citenamefont {Wang}, \citenamefont {Shi}, \citenamefont {Li}, \citenamefont
  {Fang},\ and\ \citenamefont {Bernevig}}]{song2018all}%
  \BibitemOpen
  \bibfield  {author} {\bibinfo {author} {\bibfnamefont {Z.}~\bibnamefont
  {Song}}, \bibinfo {author} {\bibfnamefont {Z.}~\bibnamefont {Wang}}, \bibinfo
  {author} {\bibfnamefont {W.}~\bibnamefont {Shi}}, \bibinfo {author}
  {\bibfnamefont {G.}~\bibnamefont {Li}}, \bibinfo {author} {\bibfnamefont
  {C.}~\bibnamefont {Fang}}, \ and\ \bibinfo {author} {\bibfnamefont {B.~A.}\
  \bibnamefont {Bernevig}},\ }\bibfield  {title} {\enquote {\bibinfo {title}
  {``all" magic angles are ``stable" topological},}\ }\href@noop {} {\bibfield
  {journal} {\bibinfo  {journal} {arXiv preprint arXiv:1807.10676}\ } (\bibinfo
  {year} {2018})}\BibitemShut {NoStop}%
\bibitem [{\citenamefont {Fu}\ and\ \citenamefont {Kane}(2006)}]{fu2006time}%
  \BibitemOpen
  \bibfield  {author} {\bibinfo {author} {\bibfnamefont {L.}~\bibnamefont
  {Fu}}\ and\ \bibinfo {author} {\bibfnamefont {C.~L.}\ \bibnamefont {Kane}},\
  }\bibfield  {title} {\enquote {\bibinfo {title} {Time reversal polarization
  and a {$Z_2$} adiabatic spin pump},}\ }\href@noop {} {\bibfield  {journal}
  {\bibinfo  {journal} {Phys. Rev. B}\ }\textbf {\bibinfo {volume} {74}},\
  \bibinfo {pages} {195312} (\bibinfo {year} {2006})}\BibitemShut {NoStop}%
\bibitem [{\citenamefont {Varjas}\ \emph {et~al.}(2015)\citenamefont {Varjas},
  \citenamefont {de~Juan},\ and\ \citenamefont {Lu}}]{varjas2015bulk}%
  \BibitemOpen
  \bibfield  {author} {\bibinfo {author} {\bibfnamefont {D.}~\bibnamefont
  {Varjas}}, \bibinfo {author} {\bibfnamefont {F.}~\bibnamefont {de~Juan}}, \
  and\ \bibinfo {author} {\bibfnamefont {Y.-M.}\ \bibnamefont {Lu}},\
  }\bibfield  {title} {\enquote {\bibinfo {title} {Bulk invariants and
  topological response in insulators and superconductors with nonsymmorphic
  symmetries},}\ }\href@noop {} {\bibfield  {journal} {\bibinfo  {journal}
  {Phys. Rev. B}\ }\textbf {\bibinfo {volume} {92}},\ \bibinfo {pages} {195116}
  (\bibinfo {year} {2015})}\BibitemShut {NoStop}%
\bibitem [{\citenamefont {Benjamin}\ and\ \citenamefont
  {Bernevig}(2018)}]{benjamin2018the}%
  \BibitemOpen
  \bibfield  {author} {\bibinfo {author} {\bibfnamefont {J.~Wieder}\
  \bibnamefont {Benjamin}}\ and\ \bibinfo {author} {\bibfnamefont {B.~Andrei}\
  \bibnamefont {Bernevig}},\ }\bibfield  {title} {\enquote {\bibinfo {title}
  {The axion insulator as a pump of fragile topology},}\ }\href@noop {}
  {\bibfield  {journal} {\bibinfo  {journal} {arXiv preprint arXiv:1810.02373}\
  } (\bibinfo {year} {2018})}\BibitemShut {NoStop}%
\bibitem [{\citenamefont {Nakahara}(2003)}]{nakahara2003geometry}%
  \BibitemOpen
  \bibfield  {author} {\bibinfo {author} {\bibfnamefont {M.}~\bibnamefont
  {Nakahara}},\ }\href@noop {} {\emph {\bibinfo {title} {Geometry, Topology and
  Physics}}}\ (\bibinfo  {publisher} {CRC Press},\ \bibinfo {year}
  {2003})\BibitemShut {NoStop}%
\bibitem [{\citenamefont {Murakami}\ \emph {et~al.}(2007)\citenamefont
  {Murakami}, \citenamefont {Iso}, \citenamefont {Avishai}, \citenamefont
  {Onoda},\ and\ \citenamefont {Nagaosa}}]{murakami2007tuning}%
  \BibitemOpen
  \bibfield  {author} {\bibinfo {author} {\bibfnamefont {S.}~\bibnamefont
  {Murakami}}, \bibinfo {author} {\bibfnamefont {S.}~\bibnamefont {Iso}},
  \bibinfo {author} {\bibfnamefont {Y.}~\bibnamefont {Avishai}}, \bibinfo
  {author} {\bibfnamefont {M.}~\bibnamefont {Onoda}}, \ and\ \bibinfo {author}
  {\bibfnamefont {N.}~\bibnamefont {Nagaosa}},\ }\bibfield  {title} {\enquote
  {\bibinfo {title} {Tuning phase transition between quantum spin hall and
  ordinary insulating phases},}\ }\href@noop {} {\bibfield  {journal} {\bibinfo
   {journal} {Phys. Rev. B}\ }\textbf {\bibinfo {volume} {76}},\ \bibinfo
  {pages} {205304} (\bibinfo {year} {2007})}\BibitemShut {NoStop}%
\end{thebibliography}


\end{document}